\newif\ifanonymous
\anonymousfalse    

\ifanonymous
\documentclass[journal=tches,submission,xcolor=table]{iacrtrans}
\else
\documentclass[journal=tches,xcolor=table]{iacrtrans}

\fi

\usepackage[nolist]{acronym}
\begin{acronym}

    \acro{3DES}{Triple-DES}
    
    \acro{ACT-R}{Adaptive Control of Thought-Rational}
    \acro{AES}{Advanced Encryption Standard}
    \acro{AI}{Artificial Intelligence}
    \acro{ALU}{Arithmetic Logic Unit}
    \acro{ANOVA}{Analysis of Variance}
    \acroplural{ANOVA}[ANOVAs]{Analyses of Variance}
    \acro{API}{Application Programming Interface}
    \acro{AOI}{Area of Interest}
    \acroplural{AOI}[AOIs]{Areas of Interest}
    \acro{ARX}{Addition Rotation XOR}
    \acro{ATPG}{Automatic Test Pattern Generation}
    \acro{ASIC}{Application Specific Integrated Circuit}
    \acro{ASIP}{Application Specific Instruction-Set Processor}
    \acro{AS}{Active Serial}
    
    \acro{BDD}{Binary Decision Diagram}
    \acro{BFS}{Breadth First Search}
    \acro{BGL}{Boost Graph Library}
    \acro{BKA}{German Federal Criminal Police Office}
    \acro{BNF}{Backus-Naur Form}
    \acro{BRAM}{Block-Ram}
    \acro{BSE}{Backscattered Electron}
    
    \acro{CBC}{Cipher Block Chaining}
    \acro{CFB}{Cipher Feedback Mode}
    \acro{CFG}{Control Flow Graph}
    \acro{CfP}{Call for Papers}
    \acroplural{CfP}[CfPs]{Calls for Papers}
    \acro{CIA}{confidentiality, integrity, and availability}
    \acro{CLB}{Configurable Logic Block}
    \acro{CLI}{Command Line Interface}
    \acro{CMOS}{Complementary Metal-Oxide-Semiconductor}
    \acro{CMP}{Chemical-Mechanical Polishing}
    \acro{CNN}{Convolutional Neural Network}
    \acro{COFF}{Common Object File Format}
    \acro{CPA}{Correlation Power Analysis}
    \acro{CPU}{Central Processing Unit}
    \acro{CRediT}{Contributor Roles Taxonomy}
    \acro{CRC}{Cyclic Redundancy Check}
    \acro{CTA}{Concurrent Think Aloud}
    \acro{CTR}{Counter}
    
    \acro{DAA}{Disassemble / Analyze / Assemble}
    \acro{DAC}{Design Automation Conference}
    \acro{DATE}{Design, Automation and Test in Europe}
    \acro{DC}{Direct Current}
    \acro{DES}{Data Encryption Standard}
    \acro{DFA}{Differential Frequency Analysis}
    \acro{DFT}{Discrete Fourier Transform}
    \acro{DIP}{Distinguishing Input Pattern}
    \acro{DL}{Deep Learning}
    \acro{DLL}{Dynamic Link Library}
    \acro{DMA}{Direct Memory Access}
    \acro{DNF}{Disjunctive Normal Form}
    \acro{DPA}{Differential Power Analysis}
    \acro{DSO}{Digital Storage Oscilloscope}
    \acro{DSP}{Digital Signal Processing}
    \acro{DUT}{Design Under Test}
    
    \acro{ECB}{Electronic Code Book}
    \acro{ECC}{Elliptic Curve Cryptography}
    \acro{EDA}{Electronic Design Automation}
    \acro{EEPROM}{Electrically Erasable Programmable Read-only Memory}
    \acro{EMA}{Electromagnetic Emanation}
    \acro{EM}{electro-magnetic}
    \acro{EMME}{Eye Movement Modeling Example}
    \acro{ESD}{Electrically Significant Differences}
    \acro{EU}{European Union}
    
    \acro{FFT}{Fast Fourier Transformation}
    \acro{FF}{Flip Flop}
    \acro{FI}{Fault Injection}
    \acro{FIR}{Finite Impulse Response}
    \acro{FPGA}{Field-Programmable Gate Array}
    \acro{FSM}{Finite State Machine}

    \acro{GDSII}{Graphic Data System II}
    \acro{GLMM}{Generalized Linear Mixed Model}
    \acro{GLM}{Generalized Linear Model}
    \acro{GNN}{Graph Neural Network}
    \acro{GT}{Grounded Theory}
    \acro{GUI}{Graphical User Interface}
    
    \acro{HCI}{Human Computer Interaction}
    \acro{HDL}{Hardware Description Language}
    \acro{HD}{Hamming Distance}
    \acro{HF}{High Frequency}
	\acro{HRE}{Hardware Reverse Engineering}
    \acro{HSM}{Hardware Security Module}
    \acro{HW}{Hamming Weight}
    
    \acro{IC}{Integrated Circuit}
    \acro{ICC}{Intraclass Correlation}
    \acro{IO}[I/O]{Input/Output}
    \acro{IOB}{Input Output Block}
    \acro{IoT}{Internet of Things}
    \acro{IRB}{Institutional Review Board}
    \acro{IP}{Intellectual Property}
    \acro{IPIP}{International Personality Item Pool}
    \acro{IPIP-NEO}{International Personality Item Pool Representation of the NEO PI-R}
    \acro{IQ}{Intelligence Quotient}
    \acro{IRR}{Inter-Rater Reliability}
    \acro{ISA}{Instruction Set Architecture}
    \acro{ISCED}{International Standard Classification of Education}
    \acro{ITS}{Intelligent Tutoring System}
    \acro{IoU}{Intersection over Union}
    \acro{IV}{Initialization Vector}
    
    \acro{JTAG}{Joint Test Action Group}
    
    \acro{KAT}{Known Answer Test}
    \acro{KLI}{Knowledge-Learning-Instruction}

    \acro{LFSR}{Linear Feedback Shift Register}
    \acro{LMM}{Linear Mixed Model}
    \acro{LSB}{Least Significant Bit}
    \acro{LUT}{Lookup table}
    
    \acro{MAC}{Message Authentication Code}
    \acro{MAD}{Median Absolute Deviation}
    \acro{MIPS}{Microprocessor without Interlocked Pipeline Stages}
    \acro{ML}{Machine Learning}
    \acro{MLM}{Multilevel Model}
    \acro{MLR}{Multiple Linear Regression}
    \acro{RMMR}{Repeated Measures Multiple Regression}
    \acro{MMIO}{Memory Mapped \acl{IO}}
    \acro{MSB}{Most Significant Bit}
    
    \acro{NASA}{National Aeronautics and Space Administration}
    \acro{NDA}{Non-Disclosure Agreement}
    \acro{NRE}{Netlist Reverse Engineering}
    \acro{NCT}{Number Connection Task}
    \acro{NSA}{National Security Agency}
    \acro{NVM}{Non-Volatile Memory}
    
    \acro{OEM}{Original Equipment Manufacturer}
    \acro{OFB}{Output Feedback Mode}
    \acro{OISC}{One Instruction Set Computer}
    \acro{ORAM}{Oblivious Random Access Memory}
    \acro{OS}{Operating System}
    \acro{OSF}{Open Science Framework}
    
    \acro{PAR}{Place-and-Route}
    \acro{PCA}{Principal Component Analysis}
    \acro{PCB}{Printed Circuit Board}
    \acro{PC}{Personal Computer}
    \acro{PCIe}{Peripheral Component Interconnect Express}
    \acro{PDK}{Process Design Kit}
    \acro{PIP}{Programmable Interconnect Point}
    \acro{PLP}{Programmable Logic Point}
    \acro{PS}{Processing Speed}
	\acro{PR}{Perceptual Reasoning}
    \acro{PUF}{Physically Unclonable Function}

    \acro{QA}{Quality Assurance}
    \acro{QBF}{Quantified Boolean Formula}

    \acro{RE}{Reverse Engineering}
    \acro{RISC}{Reduced Instruction Set Computer}
    \acro{RNG}{Random Number Generator}
    \acro{ROBDD}{Reduced Ordered Binary Decision Diagram}
    \acro{ROM}{Read-Only Memory}
    \acro{ROP}{Return-oriented Programming}
    \acro{RTA}{Retrospective Think Aloud}
    \acro{RTL}{Register Transfer Level}
    \acro{RUB}{Ruhr University Bochum}
    \acro{RFID}{Radio Frequency Identification}
    
    \acro{SAT}{Boolean Satisfiability}
    \acro{SCA}{Symbolic Computer Algebra}
    \acro{SCC}{Strongly Connected Component}
    \acro{SE}{Secondary Electron}
    \acro{SEM}{Scanning Electron Microscope}
    \acro{SEMOBS}{Self-Modifying Bitstreams}
    \acro{SHA}{Secure Hash Algorithm}
    \acro{SoK}{Systematization of Knowledge}
    \acro{SMT}{Satisfiability Modulo Theories}
    \acro{SNR}{Signal-to-Noise Ratio}
    \acro{SPA}{Simple Power Analysis}
    \acro{SPI}{Serial Peripheral Interface Bus}
    \acro{SRAM}{Static Random Access Memory}
    \acro{SRE}{Software Reverse Engineering}
    
    \acro{TA}{Think Aloud}
    \acro{TRNG}{True Random Number Generator}
    
    \acro{UART}{Universal Asynchronous Receiver Transmitter}
    \acro{UHF}{Ultra-High Frequency}
    \acro{UI}{User Interface}
    \acro{US}[U.S.]{United States}
    
    \acro{vFPGA}{virtual FPGA}
    \acro{VC}{Verbal Comprehension}
    \acro{VLSI}{Very-Large-Scale Integration}
    
    \acro{WAIS-IV}{Wechsler Adult Intelligence Scale}
    \acro{WM}{Working Memory}
    \acro{WISC}{Writeable Instruction Set Computer}
    
    \acro{XDL}{Xilinx Description Language}
    \acro{XTS}{XEX-based Tweaked-codebook with ciphertext Stealing}

    \acro{ZITiS}{German Central Office for Information Technology in the Security Sector}
    \acro{ZVT}{Zahlen-Verbindungs-Test}    
\end{acronym}

\usepackage{tikz}
\usepackage{amsmath}
\usepackage{pgfplots}
\pgfplotsset{compat=1.18}
\usepackage{csquotes}
\usepackage{booktabs}
\usepackage{enumerate}
\usepackage{enumitem}
\usepackage{pdfpages}
\usepackage{xspace}
\usepackage{comment} 
\usepackage[most]{tcolorbox} 
\usepackage[framemethod=tikz]{mdframed} 
\usepackage{multirow} 
\usepackage{pifont}
\usepackage{caption}
\usepackage{tabularx}
\usepackage{siunitx}
\usepackage{orcidlink}
\usepackage{xurl}

\usepackage{subcaption}
\usepackage{sidecap}

\newcommand{\ie}{i.\,e.}
\newcommand{\eg}{e.\,g.}

\newcommand{\etal}{\textit{et al.}\@\xspace}

\newcolumntype{Y}{>{\centering\arraybackslash}X}
\newcolumntype{L}[1]{>{\raggedright\let\newline\\\arraybackslash\hspace{0pt}}m{#1}}
\newcolumntype{C}[1]{>{\centering\let\newline\\\arraybackslash\hspace{0pt}}m{#1}}
\newcolumntype{R}[1]{>{\raggedleft\let\newline\\\arraybackslash\hspace{0pt}}m{#1}}

\newtcbox{\blackbox}[1][]{
  on line,
  colback=black,
  colframe=black,
  boxrule=0.8pt,
  boxsep=0pt,
  left=2pt, right=2pt,
  top=2pt, bottom=2pt,
  arc=1pt,
  box align=base,
  fontupper=\fontsize{7pt}{7pt}\color{white}\bfseries,
  #1
}

\newtcbox{\termbox}[1][]{
  on line,
  colback=chesblue!35,
  colframe=chesblue!95!black,
  boxrule=0.8pt,
  boxsep=0pt,
  left=2pt, right=2pt,
  top=2pt, bottom=2pt,
  arc=1pt,
  box align=base,
  fontupper=\fontsize{7pt}{7pt}\color{black}\bfseries,
  #1
}

\definecolor{chesblue}{HTML}{2B579A}

\newcommand{\proposalbox}[1]{%
  \begin{tcolorbox}[
    enhanced,
    frame hidden,
    colback=chesblue!8,
    borderline west={0.8pt}{0pt}{chesblue!85},
    sharp corners,
    before skip=6pt,
    after skip=3pt,
    boxsep=.8pt,
    left=4pt, right=2pt,
    top=1pt, bottom=1pt,
    toprule=0.01pt,
    bottomrule=0.01pt
  ]
    \textbf{Proposal:}#1
  \end{tcolorbox}%
}

\newcommand{\stakeholderProposals}[2]{%
  \begin{tcolorbox}[
    enhanced,
    breakable,
    frame hidden,
    colback=chesblue!8,
    borderline west={0.8pt}{0pt}{chesblue!85},
    sharp corners,
    before skip=6pt,
    after skip=3pt,
    boxsep=.8pt,
    left=4pt, right=2pt,
    top=1pt, bottom=1pt,
    toprule=0.01pt,
    bottomrule=0.01pt
  ]
    \textbf{#1:}#2
  \end{tcolorbox}%
}

\newmdenv[
  backgroundcolor=gray!8,
  linewidth=0.4pt,
  roundcorner=4pt,
  skipabove=6pt,
  skipbelow=0pt,
  innerleftmargin=5pt,
  innerrightmargin=5pt,
  innertopmargin=3pt,
  innerbottommargin=3pt
]{rqbox}

\newif\iftodo
\todotrue

\newif\ifrevise
\revisetrue

\newcommand{\cmark}{\textcolor{green!60!black}{\ding{51}}}  
\newcommand{\xmark}{\textcolor{red}{\ding{55}}}             

\newcommand{\CIRCLEEMPTY}{%
  \tikz[baseline=-0.5ex]{\draw (0,0) circle (0.11cm);}%
}

\newcommand{\CIRCLEHALF}{%
  \tikz[baseline=-0.5ex]{%
    \draw (0,0) circle (0.11cm);
    \fill (-0.11,0) arc (180:360:0.11cm) -- (0,0) -- cycle;
  }%
}

\newcommand{\CIRCLEFULL}{%
  \tikz[baseline=-0.6ex]{\fill (0,0) circle (0.11cm);}%
}

\newcommand{\CIRCLETHREEQUARTER}{%
  \tikz[baseline=-0.5ex]{%
    \draw (0,0) circle (0.11cm);
    \fill (-0.11,0) arc (180:360:0.11cm) -- (0,0) -- cycle;
    \fill (0,0) -- (0,0.11) arc (90:180:0.11cm) -- cycle;
  }%
}

\author[Z. Karadağ, S. Klix, R. Walendy, F. Hahn, K. Dorschel, J. Speith, C. Paar, S. Becker]{%
  Zehra Karadağ\inst{1,2}\textsuperscript{$\ast$}\orcidlink{0009-0001-6863-3184} \and
  Simon Klix\inst{2}\textsuperscript{$\ast$}\orcidlink{0000-0002-9369-2901} \and
  René Walendy\inst{2}\textsuperscript{$\ast$}\orcidlink{0000-0002-5378-3833} \and
  Felix Hahn\inst{2}\textsuperscript{$\ast$}\orcidlink{0009-0008-9260-4288} \and
  Kolja Dorschel\inst{2}\orcidlink{0009-0004-8593-2866} \and
  Julian Speith\inst{2}\orcidlink{0000-0002-8408-8518} \and
  Christof Paar\inst{2}\orcidlink{0000-0001-8681-2277} \and
  Steffen Becker\inst{1,2}\orcidlink{0000-0001-7526-5597}} 
\institute{
  Ruhr University Bochum (RUB), Bochum, Germany,\\
  \{\email[zehra.karadag@rub.de]{zehra.karadag}, \email[steffen.becker@rub.de]{steffen.becker}\}@rub.de
  \and
  Max Planck Institute for Security and Privacy (MPI-SP), Bochum, Germany,
  \{\email[simon.klix@mpi-sp.org]{simon.klix}, \email[rene.walendy@mpi-sp.org]{rene.walendy}, \email[felix.hahn@mpi-sp.org]{felix.hahn}, \email[kolja.dorschel@mpi-sp.org]{kolja.dorschel}, \email[julian.speith@mpi-sp.org]{julian.speith}, \email[christof.paar@mpi-sp.org]{christof.paar}\}@mpi-sp.org
}

\title{SoK: From Silicon to Netlist and Beyond}
\subtitle{Two Decades of Hardware Reverse Engineering Research}

\begin{document}

\maketitle

{\let\thefootnote\relax\footnotetext{\vspace{-\baselineskip}\textsuperscript{$\ast$}These authors contributed equally to this work.}}

\keywords{Hardware Reverse Engineering \and Systematization of Knowledge \and IC Reverse Engineering \and FPGA Reverse Engineering \and Netlist Reverse Engineering}

\begin{abstract}
Hardware serves as the root of trust in modern computing systems, making Hardware Reverse Engineering (HRE) essential for security assurance---from design verification and supply-chain integrity to vulnerability discovery.
We scope HRE to netlist recovery and its subsequent analysis, spanning the three subdomains of Integrated Circuit (IC), Field-Programmable Gate Array (FPGA), and netlist reverse engineering.
These subdomains differ in their methodologies, but share core processes and are shaped by common requirements and legal constraints of the same stakeholders.
Despite an increasing number of publications, the field lacks a systematic understanding of how these obstacles have stunted the research ecosystem.
To address this gap, we present the first large-scale Systematization of Knowledge (SoK) of the HRE workflow, analyzing 187 peer-reviewed publications.
Across all three subdomains, we identify eleven concrete technical challenges---from a widening gap between academic research and modern semiconductor technology nodes to overly idealized assumptions in netlist analysis---and propose actionable directions for each.
A retrospective evaluation of all 30 published artifacts reveals that key results could be reproduced for only seven, a mere 4\,\% of all 187 papers in our corpus, confirming a systemic reproducibility crisis.
We trace both the technical and reproducibility challenges to three structural barriers that recur across all subdomains: scarce reusable artifacts, missing benchmarks, and unresolved legal constraints on data sharing and collaboration.
Based on these findings, we derive stakeholder-specific recommendations for academia, industry, and government to transition HRE from isolated research silos toward a collaborative discipline capable of assuring increasingly complex, global hardware supply chains.
\end{abstract}

\acresetall
\section{Introduction}
\label{section:introduction}
Modern computing systems increasingly rely on a silicon root of trust, especially in high-assurance and safety-critical settings~\cite{DBLP:journals/tecs/CianiPMBBKPGAR25}.
Yet hardware manufacturing depends on globally distributed, complex supply chains that create opportunities for hardware Trojan insertion~\cite{Adee2008TheHF}, counterfeiting~\cite{Guin2014CounterfeitIC}, and intellectual property theft~\cite{DBLP:conf/coins/KnechtelPS19}.
These threats have driven research into \acf{HRE}, which aims to reconstruct higher levels of abstraction from fabricated hardware or intermediate design representations~\cite{chisholm1999understanding}, as a key capability for establishing trust in hardware provenance and integrity~\cite{DBLP:conf/sigcse/WalendyW0PR25}.
While other approaches also extract information from hardware devices (\eg, side-channel analysis), we restrict our definition of \ac{HRE} to methods that recover and analyze netlists to achieve high-level understanding.
In practice, \ac{HRE} enables security evaluations such as vulnerability discovery~\cite{nohl2008reverseengineering}, hardware verification and Trojan detection~\cite{puschner2023red}, and counterfeit detection and avoidance~\cite{8050605,botero2021hardware}.
Because \ac{HRE} can also be used offensively to extract design secrets~\cite{klix2024stealing} and enable hardware-level attacks~\cite{kammerstetter2014breaking,DBLP:journals/jhss/KrachenfelsLSDF20,leander2024hawkeye}, public research is equally necessary for assessing attacker capabilities and evaluating defenses.

\begin{figure}[htb]
    \centering
    \includegraphics[width=0.49\linewidth]{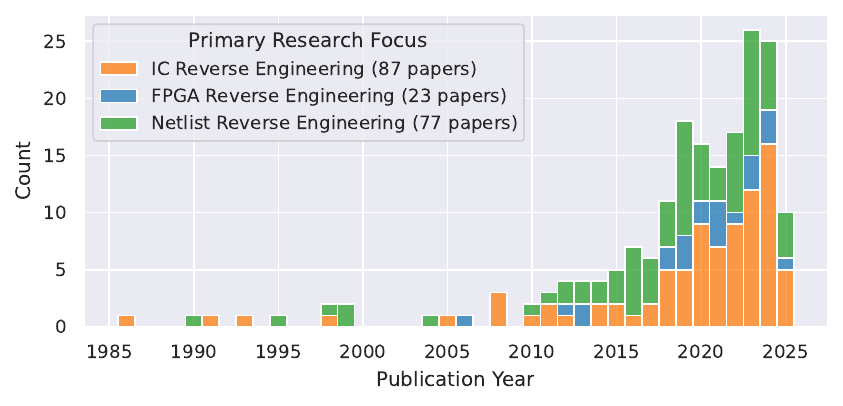}%
    \hfill
    \includegraphics[width=0.49\linewidth]{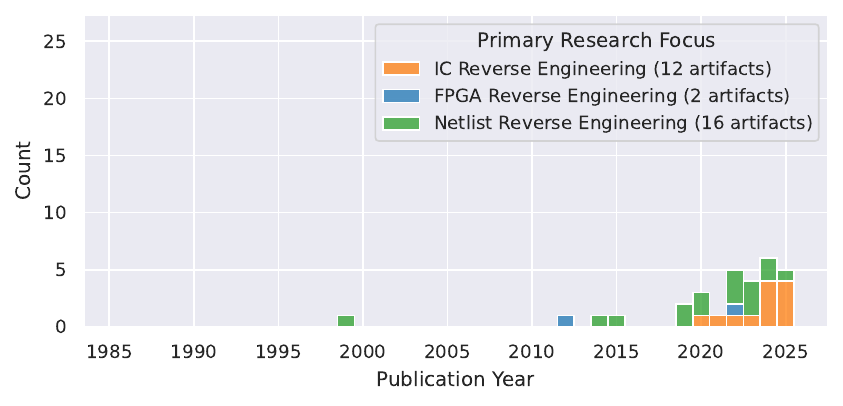}
    \caption{Annual count of peer-reviewed \acl{HRE} publications (left) and publicly available artifacts (right) in our corpus, grouped by primary research topic.}
    \label{fig:intro:pub-artifact-dist}
\end{figure}

After two decades of growth (see \autoref{fig:intro:pub-artifact-dist}), \ac{HRE} has become an established but fragmented research field. 
It spans the three subdomains of \acf{IC}, \acf{FPGA}, and netlist reverse engineering, with contributions from hardware design, manufacturing, failure analysis, and security communities. 
Despite this diversity, the subdomains serve many of the same actors---notably government agencies, defense contractors, and specialized industry---whose requirements and constraints shape \ac{HRE} research practices.
Its extreme resource demands, from skilled professionals to multi-million-dollar sample preparation and imaging equipment, concentrate practical use in national security and supply-chain assurance, while imposing strict limits on public research.
As a consequence, the same \textit{systemic challenges}---poor reproducibility and comparability, missing cumulative progress, and legal uncertainty---recur across all three subdomains, driving many individual \textit{technical challenges} and hampering research progress.
While similar issues occur elsewhere, they are particularly consequential in \ac{HRE}, which---unlike, \eg, software reverse engineering---lacks a broad hobbyist community or widespread commercial application to independently drive tooling, knowledge sharing, and incremental progress. 
Academic research is therefore the primary vehicle for advancing public \ac{HRE} capabilities, making any obstacle to its conduct a bottleneck for the field as a whole.

Earlier surveys cover individual facets of \ac{HRE}: image processing and machine learning for \ac{IC} analysis~\cite{botero2021hardware}, algorithmic methods for netlist analysis~\cite{azriel2021survey}, and \ac{HRE} countermeasures~\cite{quadir2017survey}.
While valuable, these works are limited in scope and predate much of the current literature.
More recently, Xu \etal~\cite{cryptoeprint:2026/531} examined technical approaches and algorithmic challenges specific to netlist reverse engineering.
We take a complementary perspective and identify cross-cutting challenges that span the field rather than focusing on a single stage.
To this end, we present the first \acl{SoK} of the complete \ac{HRE} workflow---from silicon to netlist and beyond---analyzing 187 peer-reviewed publications. 
Our methodology combines iterative database searches, citation mining, and multi-author screening with structured codebooks, improving on prior surveys in scope, transparency, and reproducibility. 
We make the following contributions:

\begin{itemize}
\item \textbf{A Systematic and Comprehensive Technical Survey.}
We compile, analyze, and publish the largest corpus of peer-reviewed \ac{HRE} literature to date, jointly covering \ac{IC}, \ac{FPGA}, and netlist reverse engineering. 
For each subdomain, we characterize the state of the art along the \ac{HRE} workflow, distilling techniques and research gaps. 
The resulting corpus and its categorization provide a reference for the community.

\item \textbf{Technical Challenges and Ways Forward.}
From our analysis, we identify open \textit{technical challenges} often stemming from a lack of reproducibility, resulting in limited cumulative progress across subdomains.
Based on our findings, we then propose concrete directions for individual researchers to address each challenge.

\item \textbf{A Retrospective Artifact Evaluation.}
To further investigate the presumed reproducibility crisis in \ac{HRE}, we systematically evaluate all 30 published artifacts in our corpus along the dimensions of availability, functionality, and reproducibility, following established artifact evaluation practices. 
Across the 30 artifacts, key results could be reproduced with only seven, equaling a mere 4\,\% of all 187 papers in our corpus---confirming substantial reproducibility limitations in \ac{HRE} research.
To ease their future use, we provide scripts that automate the setup of the available artifacts.

\item \textbf{Cross-Cutting Opportunities and Stakeholder Recommendations.}
We analyze \textit{systemic challenges} that recur across all \ac{HRE} subdomains and as our central synthesis identify three structural opportunities: (i)~improving reproducibility and reuse through artifact-centric practices, (ii)~enabling rigorous comparability through standardized benchmarks and evaluation metrics, and (iii)~providing legal clearance and clarity for public \ac{HRE} research.
For these opportunities, we derive concrete, stakeholder-specific recommendations for academia, industry, and government.
\end{itemize}

\noindent 
We argue that, because the same actors fund, conduct, and consume research across all three \ac{HRE} subdomains, coordinated action is not only feasible but also necessary to address structural barriers that no single community can resolve in isolation.

\paragraph{Data Availability.}
\label{par:introduction:data}
The supplementary material accompanying this paper comprises three components,
permanently archived at \url{https://doi.org/10.5281/zenodo.20797192}:
(i)~\textbf{literature-related material}, including the complete literature
corpus, the database search query, the codebook used to 
categorize papers, and all citations identified during the initial search and
subsequent citation mining rounds with their respective screening
decisions;
(ii)~\textbf{per-artifact evaluation notes}, detailing the reasoning behind
availability, functionality, and reproducibility ratings
for each of the 30 artifacts; and
(iii)~\textbf{Dockerfile and scripts}, automating the retrieval and installation of the evaluated tools and datasets.

\section{Methodology}
\label{section:methodology}
\begin{figure}[htb]
    \centering
    \includegraphics[width=\textwidth]{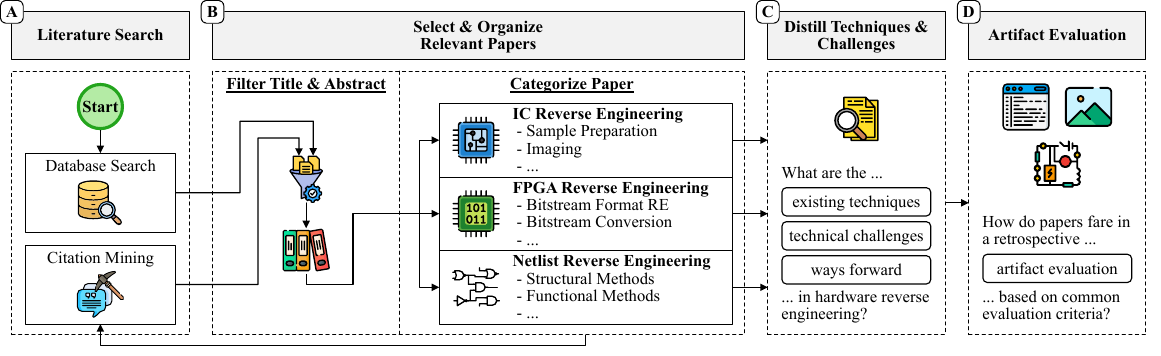}
     \caption{Four-stage pipeline for systematizing knowledge in \acl{HRE}.}
    \label{fig:methodology_flow}
\end{figure}

As illustrated in \autoref{fig:methodology_flow}, we followed a structured four-stage analysis pipeline, guided by the methodology of Wolfswinkel~\etal~\cite{wolfswinkel2013grounded}.
Starting with database searches (Step~A), we filter and organize the discovered papers (Step~B).
The retained papers are then used as a seed set to iteratively expand the corpus through citation mining, with newly identified papers subjected to the same filtering process (Step~B).
We then analyze the resulting papers in detail (Step~C) and examine the available artifacts according to the common artifact badge categories \textit{Availability}, \textit{Functionality}, and \textit{Reproducibility} (Step~D).

\subsection{Analysis Pipeline}

\paragraph{Step A: Literature Search.} 
We identified 571 papers through database searches of DBLP, Semantic Scholar, and Google Scholar using a search query that was designed based on domain expertise and highly cited \ac{HRE} publications~\cite{fyrbiak2017hardware,yu2016formal,botero2021hardware}.
To ensure comprehensive coverage of \ac{HRE} literature, we performed backward and forward citation mining using Crossref and Semantic Scholar, including papers referenced by or referencing at least three papers in our corpus.
This step added 402 publications over four iterations, bringing the total to 973 papers.
The final iteration in December 2025 added only one paper, indicating saturation.

\paragraph{Step B: Select \& Organize Relevant Papers.} 
In each iteration, we filtered papers for relevance in two stages.
First, to determine whether a paper contributed to \ac{HRE}, one author screened each title and abstract and proposed an initial decision, which was then discussed and finalized with two additional authors.
After the final iteration, this process had resulted in 344 publications for subsequent categorization.
Second, two authors reviewed each paper and labeled it according to the topics it covers, following our codebook. 
For each paper, exactly one primary code and any number of secondary codes were assigned.
Additionally, each paper received at least one type label based on its primary contribution.
Disagreements were resolved in discussions with a third author.

In both stages, papers were excluded if they were not written in English, not peer-reviewed, behind paywalls despite standard institutional subscriptions, or written as technical reports, dissertations, or tutorials.
Papers were also excluded if they did not primarily contribute to netlist recovery or netlist analysis.
Examples of such exclusions encompass research on pedagogical aspects, hardware design, side-channel analysis, Trojans, and obfuscation.
While Trojan and obfuscation research may overlap with netlist analysis, they typically investigate niche challenges, such as removing a specific obfuscation or detecting a particular Trojan.
Therefore, we consider them out of scope.
Ultimately, this process yielded a final corpus of 187 papers, divided into the three \ac{HRE} subdomains: 87 focused on \ac{IC}, 23 on \ac{FPGA}, and 77 on netlist reverse engineering.

\paragraph{Step C: Distill Techniques and Propose Pathways Forward.} 
Three two-person teams---one per subdomain---analyzed all publications.
Each researcher read a subset of papers, noting methods, challenges, and evaluation procedures, to identify key techniques and open challenges in each \ac{HRE} subdomain.
We then distilled subdomain-specific and cross-cutting proposals tailored to academia, industry, and government.

\paragraph{Step D: Artifact Evaluation.}
We categorized all \ac{HRE} artifacts as tools, images, or other resources (\eg, scripts or \ac{RTL} code) and evaluated them for \textit{Availability}, \textit{Functionality}, and \textit{Reproducibility}.
Availability ranges from permanent repositories (\eg, Zenodo), non-permanent hosting (\eg, GitHub or personal websites), availability upon request, and unavailable artifacts (no longer hosted or no response to requests).
Functionality comprises \textit{completeness} (all claimed components are provided), \textit{documentation quality}, and \textit{exercisability} (successful execution)---the latter assessed only for tools.
Reproducibility was assessed by whether key results in the paper  could be reproduced.
Two research assistants with prior \ac{HRE} experience conducted the evaluations and documented their findings using a structured template.
Each artifact received a minimum of six hours of investigation, often substantially more, before concluding it could not be executed or did not reproduce key results.
The process was supported by weekly meetings with two authors over more than six months, and each assessment was independently verified by one author.
The analysis was conducted without contacting the artifacts' creators:
All available tools were executable, and reproduction failures stemmed from missing materials (binaries, datasets, ground truth, scoring tools) rather than from issues that consulting the creators could have resolved.

\subsection{Limitations}
Our literature search may have excluded relevant work written in other languages.
Additionally, our exclusion criteria may have omitted papers containing relevant insights near the boundaries of our scope.
These trade-offs were necessary to maintain feasibility, and we argue that their impact was mitigated through iterative backward and forward citation mining until saturation.
Furthermore, since parts of the screening relied on individual judgment, alternative decisions could yield slightly different results.
To reduce such effects, all decisions were reviewed through multi-author discussions and consensus procedures.

\section{\texorpdfstring{\ac{HRE}}{HRE} Techniques, Challenges, and Pathways Forward}
\label{section:results}

\begin{figure}[htb]
    \centering
    \includegraphics[width=0.7\linewidth]{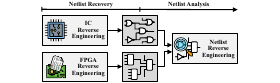}
    \caption{High-level overview of the end-to-end reverse engineering process.
    Reconstruction of a gate-level netlist from an \ac{IC} or \ac{FPGA}, followed by the analysis of the recovered netlist.}
    \label{hre_sok::figure::hre_flow}
\end{figure}

As depicted in \autoref{hre_sok::figure::hre_flow}, \ac{HRE} comprises two phases: netlist recovery and analysis~\cite{azriel2021survey} 
with recovery differing between \acp{IC} and \acp{FPGA}~\cite{wallat2019highway}.
For each subdomain, we discuss state-of-the-art techniques, open technical challenges, and how to address them.

\subsection{From \texorpdfstring{\ac{IC}}{IC} to Netlist}
Modern digital \acp{IC} consist of large-scale Boolean logic circuits, realized by hundreds of thousands to billions~\cite{hanindhito2025technology} of nanoscopic transistors on the polysilicon layer.
Stacked metal layers connect these transistors to form a circuit. 
Extracting the netlist involves recovering both transistors and interconnects, generally in a destructive manner, and relies heavily on advanced imaging, computer vision, and machine learning techniques.

\subsubsection{Techniques for \texorpdfstring{\ac{IC}}{IC} Reverse Engineering}
\begin{figure}[htb]
    \centering
    \includegraphics[width=0.7\linewidth]{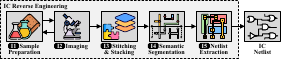}
    \caption{\ac{IC} reverse engineering flow.}
    \label{hre_sok::figure::ic_flow}
\end{figure}

The literature generally describes the recovery of a netlist from an \ac{IC} as a five-stage process (see \autoref{hre_sok::figure::ic_flow}): \blackbox{I1} sample preparation, \blackbox{I2} imaging, \blackbox{I3} stitching and stacking, \blackbox{I4} semantic segmentation, and \blackbox{I5} netlist extraction.

\paragraph{\blackbox{I1} Sample Preparation.} 
An \ac{IC} is protected by an encapsulation, also known as a package.
The package ensures physical durability~\cite{blythe1993layout} and provides connections, \eg, to a circuit board~\cite{greig2007integrated}.
This package must be removed to access the die before the delayering process can reveal the circuit.

\textit{Depackaging.}
Depackaging methods depend on the package material.
Packages are typically made of plastic, metal, or ceramic~\cite{malcik2011microscopic, blythe1993layout}.
Plastic packages constitute the majority and are primarily removed using acid~\cite{maitra2024physical, scholl2021dielectric, kimura2020decomposition}.
The literature provides information on acid mixture ratios and processing temperatures required for effective removal~\cite{malcik2011microscopic, kammerstetter2014breaking}.
Some packages allow for the top cover to be removed mechanically instead~\cite{maitra2023microstructural}.
Advanced packages may encapsulate multiple dies under a single metal lid, which can be removed by milling~\cite{padro2023assessment}.
Decapsulation of ceramic packages has not been discussed in our corpus of \ac{HRE} literature.

\textit{Cross-Section.}
A first step in analyzing a die is often sacrificing one sample by cutting across all of its layers.
The resulting cross-section reveals the materials and thickness of each layer~\cite{lippmann2025multipartner, lippmann2019integrated, padro2023assessment}, guiding the choice of delayering methods.

\textit{Delayering.}
Layer-removal methodologies can be classified into several categories.
\textit{Wet etching} removes material through chemical reactions in liquid solutions~\cite{lippmann2019integrated,maitra2023microstructural,kimura2020decomposition}.
\textit{Mechanical processing} relies on direct mechanical contact with a solid tool or abrasive~\cite{nohl2008reverseengineering,kammerstetter2014breaking,lippmann2019integrated,scholl2021dielectric, padro2024quantitative,lippmann2020verification}.
It may be combined with chemistry, as in chemical-mechanical polishing~\cite{padro2023assessment}.
\textit{Dry etching} comprises a variety of techniques in the gas phase or vacuum using ions, plasma, or reactive gases~\cite{lippmann2019integrated, kimura2020decomposition}.
For instance, in \textit{ion milling}, a beam of energetic ions is accelerated toward the sample, physically sputtering atoms from its surface~\cite{rothaug2025advancing, rothaug2023unsupervised, maitra2024physical}.
Ion milling can be chemically assisted~\cite{waite2021preparation}; however, material removal remains ion-driven.
It is not to be confused with \textit{reactive ion etching}, which also uses ion bombardment but removes material primarily chemically~\cite{principe2017steps,kim2018fast}.

During the delayering process, it is important to keep the surface planar~\cite{kammerstetter2014breaking, courbon2020practical,bette2022automated,scholl2020sample}, and to know when to stop to avoid damage to subsequent layers, a process called endpoint detection.
The endpoint can be detected using manual \ac{SEM} inspection~\cite{lippmann2019integrated}, or other methods such as ultraviolet photon spectroscopy emission~\cite{dibattista2024large}, or optical microscopy~\cite{scholl2021dielectric, lippmann2020verification, padro2024quantitative}.

\paragraph{\blackbox{I2} Imaging.}
Imaging \acp{IC} is usually done with an \ac{SEM}.
Proper use requires balancing parameters to obtain reliable results in a reasonable time, and improper settings can produce low-quality images.
Optimal settings vary depending on the instrument, even when imaging the same layer of the same \ac{IC} \cite{ludwig2021enabling}.
Settings and detector choice also depend on the type of layer to be imaged.
While detectors can be combined~\cite {lippmann2024expert}, they are typically used individually.
For metal layers, backscattered electron detectors are preferred, as they provide the necessary contrast to distinguish metal from surrounding insulation \cite{rothaug2023unsupervised, kimura2020decomposition}. 
For imaging transistor gates, however, secondary electron detectors are used to reveal fine surface topography~\cite{blythe1993layout, bette2023noreference}.

\paragraph{\blackbox{I3} Stitching \& Stacking.}
Since the field of view of an \ac{SEM} is limited, it cannot capture an entire layer at once.
Consequently, each layer is scanned in a two-dimensional grid, with every image depicting only a small tile of the layer.
To reconstruct transitions between images, tiles need to be \textit{stitched} together.
To facilitate this stitching, images are taken with an overlap.
Although some modern imaging systems offer automated stitching, researchers also use custom algorithms~\cite{torrance2011stateoftheart} or adapt tools from other domains~\cite{nohl2008reverseengineering, kammerstetter2014breaking}.
To this end, adjacent tiles are first positioned locally, \ie, relative to each other, leveraging overlap using techniques such as phase correlation~\cite{lin2020deep, singla2021recovery}, normalized cross-correlation~\cite{singla2021recovery}, or feature extraction~\cite{singla2021recovery, lippmann2019integrated}.
Because local stitching can be imperfect, global stitching optimizes the mosaic by considering multiple neighboring tiles, either sequentially~\cite{burian2022automated} or simultaneously~\cite{lippmann2020verification,singla2021recovery}.
To recover signal routing across layers and reveal the \ac{IC}'s 3D structure, the stitched layers must be \textit{stacked}.
To this end, anchor points are defined between adjacent layers, and the layer image is warped accordingly~\cite{quijada2018largearea}.
We found no use of deep learning for stitching, though it has been applied to stack alignment~\cite{lin2020deep,singla2021recovery} and stitching error detection~\cite{lin2020deep}.

\paragraph{\blackbox{I4} Semantic Segmentation.} 
Wires, vias, transistor gates, and active regions are identified through semantic segmentation, with optional pre- and post-processing.

\textit{Pre-Processing.}
\ac{IC} images can be pre-processed to facilitate subsequent recognition methods, although these approaches do not always improve results~\cite{machadotrindade2018segmentation}.
This step typically has two objectives, which are often achieved using classical computer vision methods:
First, noise is suppressed through averaging of multiple images~\cite{lippmann2019integrated}, or spatial filters incorporating surrounding pixels~\cite{masalskis2008reverse,machadotrindade2018segmentation}.
Second, adjusting brightness and contrast corrects for visual inconsistencies between tiles caused by acquisition variations~\cite{tee2023unsupervised,machadotrindade2018segmentation}.
Deep learning has also been applied for noise reduction~\cite{xiao2024tadensenet}, but requires retraining for each new dataset~\cite{giannatou2019deep}.

\textit{Segmentation.}
To eventually reconstruct a gate-level netlist, all visible circuit elements on each layer must be identified.
Historically, this identification was performed manually by crawling over printed images~\cite{torrance2011stateoftheart}, and later using digital annotation software~\cite{quijada2014use,kammerstetter2014breaking}.
However, with the rapid increase in transistor counts, manual analysis has become impractical, prompting the emergence of automated methods.
For metal layers, researchers first relied on classical computer vision methods that group pixels based on intensity thresholds~\cite{blythe1993layout,masalskis2008reverse,masalskis2010integrated} and texture~\cite{wilson2020lasre}, or detect boundaries through brightness gradients~\cite{lippmann2024expert}.
More sophisticated techniques take design rules, such as track thickness or spacing, into account~\cite{lippmann2024expert}.

An established classical approach for detecting standard cells on the polysilicon and active layer is template matching.
Here, image regions are compared against a library of visual reference patterns~\cite{courbon2015combining,quijada2014use}.
If details about the manufacturing process are unknown, a standard cell library might first need to be extracted from the images, either manually~\cite{nohl2008reverseengineering} or automatically through finding common patterns~\cite{zhu2025genetic,liu2024novel}.
Other research adopts techniques used for metal track segmentation to automatically detect standard cells by individually identifying transistors and their interconnections~\cite{huang2023gracer,quijada2018largearea}.

Recently, the focus has gradually shifted to machine-learning approaches~\cite{hong2018deep}.
The plethora of such research utilizes supervised learning~\cite{lippmann2019integrated,lin2023sem2gds,cheng2025segmentation,yang2024circuit,liu2024novel,vanderlinden2025simple} on all \ac{IC} layers; however, these models require massive amounts of labeled training data.
Manually labeling real \ac{IC} images is time-consuming~\cite{cheng2018hybrid,quijada2014use}.
To bridge this gap, standard data augmentation---such as random rotations and intensity shifts---is used to artificially expand limited datasets~\cite{yu2022datadriven,tee2023strategic}.
Beyond simple augmentation, researchers have investigated generating synthetic images from layout masks, \ie, \ac{GDSII} files.
These provide a ground truth and enable training at scale \cite{wilson2021refics, maitra2024physical,xiao2024denoising,li2024unpaired}.
Despite promising results with synthetic data~\cite{cheng2025segmentation}, domain transfer of a trained model from one \ac{IC} to another is limited, as \acp{IC} vary significantly in appearance.
Consequently, we observe a push toward unsupervised learning techniques~\cite{rothaug2025advancing} or automatic pseudo-label generation~\cite{cheng2022delayered}.
Recent publications propose using transfer learning.
To this end, Meta's SAM has been fine-tuned on \ac{IC} images~\cite{ng2024samic}.
Another technique relies on \acp{CNN} pretrained on the generic ImageNet dataset to extract general visual features from \ac{IC} images without fine-tuning~\cite{tee2023unsupervised}.
Another approach, in cases when no ground truth is available, is to transform other images with existing ground truth to resemble the unlabeled images.
A model trained on these images can then be applied to the unlabeled data~\cite{tee2022unsupervised,cheng2025segmentation}.

\textit{Post-Processing.} Post-processing can be applied to identify or reduce segmentation errors.
This prevents misclassifications from propagating into semantic violations in the extracted netlist, such as unintended open or short circuits.
Approaches from the literature correct segmentations based on neighboring pixel values, also resulting in smoother boundaries and less noise~\cite{cheng2025segmentation, yu2022datadriven, doudkin2005objects}.
Other post-processing procedures check for design rule violations related to, \eg, via size~\cite{masalskis2008reverse}, wire width~\cite{lippmann2020verification}, or spacing~\cite{masalskis2008reverse}.
Another approach detects segmentation errors using \acp{CNN}~\cite{zhang2023automatic}.

\paragraph{\blackbox{I5} Netlist Extraction.}
After all images are stitched, stacked, and segmented, a gate-level netlist can be extracted, abstracting the physical layout into a network of Boolean logic gates and sequential elements.
The connection between logic gates can be reconstructed by tracing vias and wires~\cite{nohl2008reverseengineering,lippmann2019integrated,huang2023gracer}.
The extraction mechanism for logic gates depends on the segmentation technique used.
When standard cells were identified using template matching, the vias connecting each cell to the metal layer above can be immediately identified~\cite{lippmann2019integrated}.
If transistor-level segmentation was used instead, the logic function realized by the transistors in each cell must be identified~\cite{huang2023gracer,dura2017fast,quijada2018largearea}.
To this end, subgraph matching is commonly used~\cite{huang2023gracer,putz2023plane}.

\subsubsection{Technical Challenges in \texorpdfstring{\acs{IC}}{IC} Reverse Engineering and Ways Forward}

\begin{figure}[htb]
    \centering
    \includegraphics[width=\linewidth]{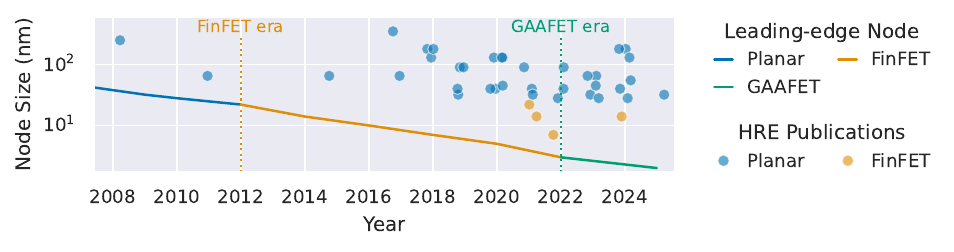}    
    \caption{Smallest node sizes across 40 publications on \ac{IC} reverse engineering, compared to the leading-edge production node~\cite{wikichip2025technology,intel2011finfet,samsung2022gaafet}. 39 additional papers did not disclose the nodes they addressed. Legacy studies ($> \SI{500}{nm}$) were omitted for clarity.}
    \label{fig:results:asic:nodes}
\end{figure}

\paragraph{Major Research---Industry Gap in Technology Node.}
The physical challenges of \ac{IC} reverse engineering are primarily driven by advancements in semiconductor manufacturing~\cite{torrance2011stateoftheart,botero2021hardware}, and academic capabilities have not kept pace~\cite{putz2023plane,lippmann2025multipartner}.
As \autoref{fig:results:asic:nodes} shows, publications predominantly consider planar CMOS nodes ranging from \SI{180}{nm} to \SI{28}{nm}, introduced between 1999 and 2010.
Only four publications address FinFET nodes (\SI{22}{nm} or below, introduced around 2012~\cite{intel2011finfet}), and none target GAAFET technology (introduced around 2022~\cite{samsung2022gaafet}), even though manufacturing has already moved to this generation.
This lag has serious implications for supply-chain security, as current nodes cannot be properly verified~\cite{botero2021hardware}.

\proposalbox{
Investigate how techniques developed for older semiconductor nodes transfer to modern technologies, identifying what can be reused, adapted, or must be redeveloped.
}

\paragraph{Error Propagation.}
The success of extracting a netlist from an \ac{IC} depends on image recognition algorithms and on the quality of images themselves~\cite{botero2021hardware,ludwig2021enabling,hong2018deep}.
Contamination such as dust particles can obscure critical circuit features~\cite{courbon2020practical,principe2017steps} and makes even manual annotation difficult \cite{tee2024integrated}. 
During delayering, non-uniform milling or etching can leave residue, causing brightness variations and geometric distortions \cite{lippmann2024expert,maitra2024physical} or, conversely, fully erode parts of the layer~\cite{padro2024quantitative}.
Furthermore, suboptimal sample preparation affects the quality of \ac{SEM} imaging.
For instance, residue can exacerbate charge-up effects \cite{padro2024quantitative,lippmann2024expert,hong2018deep}.
During imaging itself, poor parameter choice, \eg, for acceleration voltage and dwell time, can equally deteriorate image quality~\cite{lippmann2024expert,marazzi2024hifidram}.
Finally, most contemporary semantic segmentation algorithms assume error-free inputs, which is impossible to achieve in practice~\cite{botero2021hardware}.
Certain circuit structures, combined with poor image quality, also pose challenges.
Thin, closely spaced, or rough metal lines are difficult to distinguish, causing classification errors~\cite{tee2024integrated}.
Perfect results require multiple samples of the same \ac{IC}~\cite{botero2021hardware} due to step interdependencies and the risk of irreversible sample preparation errors~\cite{quijada2018largearea}.

\proposalbox{
Develop a taxonomy that classifies errors by type, prevention cost, correction cost, and downstream impact.
Crucially, this requires bridging \ac{IC} and netlist reverse engineering: some errors may be tolerable or more efficiently corrected at the netlist level (\eg, through structural consistency checks) than on \ac{SEM} images.
}

\paragraph{Limited Generalizability of Segmentation Models.}
Even for ideal images, recognition methods struggle to generalize across technology nodes and \ac{IC} layers~\cite{wilson2021refics}.
This domain shift problem~\cite{tee2022unsupervised,cheng2025unsupervised} stems from two primary sources:
First, the shapes and materials used in \acp{IC} vary significantly across nodes~\cite{wilson2022refics}.
Second, sample preparation and imaging cause variations in resolution, magnification, brightness, contrast, and noise~\cite{lippmann2020verification,wilson2021refics,wilson2020lasre,machadotrindade2018segmentation}.
Obtaining a sufficiently diverse dataset is impractical for individual research groups~\cite{tee2022unsupervised, yang2023srpyolox}, hampering the development of more robust methods.
These generalization challenges necessitate continuous parameter tuning or retraining, affecting both classical computer vision and machine learning~\cite{lin2020deep, machadotrindade2018segmentation, lippmann2020verification,cheng2018hybrid}.
While researchers are exploring solutions such as transfer learning~\cite{lin2020deep, tee2022hybrid}, automated \ac{IC} recognition across diverse conditions remains unsolved.
Generalization cannot be solved without shared, diverse datasets---yet real \ac{SEM} images are rarely published, often due to legal and \ac{IP} constraints.

\proposalbox{
Take \ac{IC} images of open-source designs manufactured using open-source \acp{PDK}, apply different preparation and imaging techniques, and publish these images with corresponding ground truth.
Only this combination of open design and \ac{PDK} enables legally compliant publication.
}

\paragraph{Image Analysis and Validation.}
The design and validation of recognition algorithms depend on accurate ground-truth labels and appropriate metrics, raising two concerns.
First, given the prevalent lack of a reliable ground truth for many datasets, manual image labeling is often required, which is arduous and costly~\cite{wilson2023secure,wilson2021refics}, particularly for large \acp{IC}~\cite{cheng2019hierarchical}, and prone to human error~\cite{vanderlinden2025simple}.
Second, comparisons of an algorithm's output with a ground truth often use pixel-accuracy metrics such as \ac{IoU}~\cite{hong2018deep, wilson2022refics}, which do not fully reflect semantic accuracy~\cite{quijada2018largearea}.
\ac{HRE}-specific metrics such as \ac{ESD} have been proposed~\cite{machadotrindade2018segmentation}, but are not widely adopted.
Still, even after addressing such concerns, fully automated recognition will remain unrealistic ~\cite{lippmann2020verification,quijada2018largearea}, making manual correction of segmentation outputs unavoidable in practice~\cite{yu2022datadriven}.

\proposalbox{
Our previous proposal would supply accurate ground-truth labels, and adopting the existing \ac{ESD} metric is straightforward.
To efficiently correct errors,
approaches integrating automated segmentation with human-in-the-loop corrections should be investigated, \eg, algorithms that output confidence scores and request manual review where needed.
}

\subsection{From \texorpdfstring{\ac{FPGA}}{FPGA} to Netlist}
\acp{FPGA} are reconfigurable hardware devices that can be programmed to implement almost any digital circuit.
To achieve this, a bitstream file in a proprietary format is loaded into the \ac{FPGA}'s configuration memory, typically implemented using \ac{SRAM}.
Because this memory is volatile, the \ac{FPGA} must be re-programmed on every power-up, so the bitstream must be stored in an external non-volatile memory.
To recover a gate-level netlist from an \ac{FPGA} bitstream, one must understand its proprietary format, recover the bitstream, potentially decrypt it, and convert it into a netlist~\cite{wallat2019highway}.

\subsubsection{Techniques for \texorpdfstring{\acs{FPGA}}{FPGA} Reverse Engineering}
\begin{figure}[htb]
    \centering
    \includegraphics[width=0.7\linewidth]{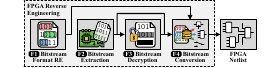}
    \caption{\ac{FPGA} reverse engineering flow.}
    \label{hre_sok::figure::fpga_flow}
\end{figure}

Based on the literature, we split \ac{FPGA} reverse engineering into four major steps (see \autoref{hre_sok::figure::fpga_flow}): \blackbox{F1}~Bitstream Format Reverse Engineering, \blackbox{F2}~Bitstream Extraction, \blackbox{F3}~Bitstream Decryption, and \blackbox{F4}~Bitstream Conversion.
Our analysis is primarily grounded in the 23 \ac{FPGA} papers included in our final corpus.
However, because the number of \ac{FPGA} reverse engineering publications is relatively small, we also extracted relevant information from work initially excluded---\eg, on bitstream protections and manipulation.

\paragraph{\blackbox{F1} Bitstream Format Reverse Engineering.}
Bitstreams follow a standard binary file structure, comprising metadata, control commands, cryptographic material, checksums, and the configuration data that defines the functionality of the programmable \ac{FPGA} fabric.
To reverse-engineer a bitstream's high-level structure, prior work has mostly relied on public documentation from \ac{FPGA} vendors~\cite{ziener2006identifying, yu2021multi, narwariya2024revbit}.
Undocumented structure information can be reverse-engineered through manual analysis using \ac{EDA} tool features such as debug bitstreams~\cite{kashani2022bitfiltrator}, or by reverse-engineering vendor software~\cite{swierczynski2015physical, moradi2011vulnerability}.

The configuration data encodes both the functionality of logic elements, aka. \acp{PLP}, such as \acp{LUT}, \acp{FF}, or hard \ac{IP} like \acp{DSP} blocks, BRAMs, and PCIe interfaces, as well as the routing between them via \acp{PIP}~\cite{ding2013deriving}.
Vendors typically do not disclose the mapping from configuration bits in the bitstream to the corresponding physical components in the \ac{FPGA} fabric~\cite{swierczynski2015fpga}.
To reveal this proprietary relationship, researchers often construct hardware designs using vendor-provided \ac{EDA} tools, generate corresponding bitstreams, and then systematically modify the positions and functions of individual programmable elements.
By comparing bitstream differences before and after each change, they identify which bits configure specific logic or routing elements~\cite{zhang2019comprehensive, benz2012bil, kim2021bitstream}.
While the core methodology remains consistent across studies, studies vary in how they improve accuracy and reduce computational effort.
\autoref{table::fpga_re::results} provides an overview of the different approaches.

Initially, a correlation-based approach was proposed, in which designs are synthesized, and the behavior of individual bits across multiple bitstreams is correlated with the resources used by each design~\cite{benz2012bil}. 
This approach was later improved by generating bitstreams that contain isolated logic or routing elements, enabling a direct one-to-one mapping between each configurable element and the bits that control it~\cite{choi2020fast}.
To reduce computational effort, some studies have leveraged the regular structure of many \ac{FPGA} architectures rather than explicitly mapping every configuration bit~\cite{zhang2019comprehensive, benz2012bil}.
Because similar elements often share a consistent configuration pattern in the bitstream, reverse engineering such patterns once can suffice to extrapolate them to other instances across the device~\cite{yu2021multi}.
The format reverse engineering process may be further sped up by reducing the number of generated bitstreams---either by partitioning bitstreams to extract information about multiple elements simultaneously~\cite{zhang2023bitfree}, or by combining information from different bitstreams~\cite{choi2020fast}.

\begin{table}[!t]
    \caption{Approaches used in step \blackbox{F1}~Bitstream Format Reverse Engineering, highlighting methodological nuances, reliance on (deprecated) AMD tooling, and reporting accuracy claims often unsubstantiated by the underlying benchmark designs.}
    \begingroup
    \scriptsize 
    \setlength{\tabcolsep}{2.3pt} 
    \begin{tabular}{p{0.09\textwidth}cccccp{0.07\textwidth}lp{0.15\textwidth}p{0.2\textwidth}p{0.21\textwidth}}
        \toprule
        & \textbf{L} & \textbf{R} & \textbf{B} & \textbf{P} & \textbf{C} & \textbf{Vendor} & \textbf{Tool} & \textbf{Device Family} & \textbf{Benchmark} & \textbf{Claimed Accuracy*} \\
        \midrule
        \rowcolor{gray!20}\cite{benz2012bil} & \CIRCLEEMPTY & \CIRCLEHALF & \xmark & \cmark & \cmark & AMD & ISE & Virtex-5 & "medium-sized design" & 80\% INT tile \acp{PIP}, \newline 14\% HCLK tile \acp{PIP} \\
        \cite{ding2013deriving} & \CIRCLEFULL & \CIRCLEFULL & \xmark & \cmark & \cmark & AMD & ISE & Spartan-II to \newline Virtex-5 & JPEG encoding sys., \ac{DSP} algorithms & 91-97\% \acp{PLP}, \newline 89-96\% \acp{PIP} \\
        \rowcolor{gray!20}\cite{swierczynski2015fpga} & \CIRCLEHALF & \CIRCLEEMPTY & \cmark & \xmark & \xmark & n.s.** & n.s. & n.s. & DES and AES S-boxes & 100\% \\ 
        \cite{jeong2018extract} & \CIRCLEHALF & \CIRCLEEMPTY & \cmark & \xmark & \xmark & AMD & Vivado & Artix-7 & resynth. one full adder & 100\% (implied) \\
        \rowcolor{gray!20}\cite{zhang2019comprehensive} & \CIRCLEFULL & \CIRCLEFULL & \xmark & \cmark & \xmark & AMD & ISE & Spartan-6 & ISCAS'85/'89, 8051 core, 68HC08, AES & 100\% \\
        \cite{leroux2019parsing} & \CIRCLEHALF & \CIRCLEEMPTY & \cmark & \xmark & \xmark & AMD & ISE & Virtex-5 & manipulation of 8 \ac{LUT} INIT values & 100\% (implied) \\
        \rowcolor{gray!20}\cite{choi2020reverse} & \CIRCLEFULL & \CIRCLEFULL & \cmark & \xmark & \xmark & AMD & ISE & Spartan-3 & 64-bit \acs{LFSR} & 88\% \\
        \cite{choi2020fast} & \CIRCLEHALF & \CIRCLEEMPTY & \cmark & \xmark & \xmark & AMD & ISE & Spartan-3, Virtex-5 & ISCAS'85, AES, DES & 100\% (implied) \\
        \rowcolor{gray!20}\cite{yu2021multi} & \CIRCLEHALF & \CIRCLEHALF & \cmark & \cmark & \xmark & AMD & Vivado & Artix-7 & 3-bit adder & 100\% \\
        \cite{kim2021bitstream} & \CIRCLEHALF & \CIRCLEHALF & \xmark & \cmark & \xmark & Mirco- \newline semi & n.s. & ProASIC3, \newline IGLOO2, Fusion & 16-bit \acs{LFSR} & 100\% \\
        \rowcolor{gray!20}\cite{danesh2021turning} & \CIRCLEFULL & \CIRCLEFULL & \xmark & \cmark & \xmark & AMD & ISE & Virtex-5 & 16 NAND based trigger circuit & 100\% recovery, 70\% interconnect stitching \\
        \cite{kashani2022bitfiltrator} & \CIRCLEHALF & \CIRCLEEMPTY & \cmark & \cmark & \xmark & AMD & Vivado & US(+) & All \acp{LUT}, All BRAMs & 100\% \\
        \rowcolor{gray!20}\cite{perumalla2023fast} & \CIRCLEHALF & \CIRCLEEMPTY & \cmark & \xmark & \xmark & AMD/ \newline Intel & n.s. & 7 Series \newline Cyclone IV/V & All \acp{LUT} & 100\% \\
        \cite{zhang2023bitfree} & \CIRCLEHALF & \CIRCLEFULL & \xmark & \cmark & \xmark & AMD & ISE & Virtex-5 \newline Virtex-7 & AES, Simon, Present ciphers & 100\% \\
        \rowcolor{gray!20}\cite{narwariya2024revbit} & \CIRCLEHALF & \CIRCLEEMPTY & \cmark & \xmark & \xmark & AMD & Vivado & 7 Series, US(+) & Custom 2, 3, and 4-bit \ac{LUT} circuits & 100\% \acp{LUT}, \newline 90+\% func. acc. \\
        \cite{narwariya2025revbit} & \CIRCLEHALF & \CIRCLEEMPTY & \cmark & \xmark & \xmark & AMD & Vivado & 7 Series, US(+) & Custom 2, 3, and 4-bit \ac{LUT} circuits & $\approx$100\% \acp{LUT}, 92\% func. acc., 88\% pin comb. \\
        \bottomrule
    \end{tabular}
    \begin{minipage}{\linewidth}
        \vspace{2pt}
        \textbf{L}ogic, \textbf{R}outing; \CIRCLEEMPTY~not addressed, \CIRCLEHALF~partially addressed (\eg, only a subset of elements recovered), \CIRCLEFULL~mapping of all elements recovered \\[2pt]
        \textbf{B}rute-force bit mappings of configurable elements by instantiating them individually, Regular-\textbf{P}attern used to infer bit mappings, \textbf{C}orrelation of configured elements with bitstreams \\[2pt]
        *Claimed accuracy refers to the reported benchmark **not specified
    \end{minipage}
    \endgroup
    \label{table::fpga_re::results}
\normalsize
\end{table}

\paragraph{\blackbox{F2} Bitstream Extraction.}
Most papers treat bitstream extraction merely as a technical prerequisite, achieved either by tapping memory buses that connect non-volatile memory to the \ac{FPGA}~\cite{yu2021multi, narwariya2025revbit} or by dumping the memory directly~\cite{fyrbiak2017hardware}.
Niche methods can extract the bitstream via software analysis if it was loaded by a co-processor~\cite{klix2024stealing} or read it back through (hidden) debug mechanisms~\cite{skorobogatov2012breakthrough, ziener2006identifying}.
Overall, bitstream extraction receives little academic attention.

\paragraph{\blackbox{F3} Bitstream Decryption.}
Most modern \acp{FPGA} provide encryption and authentication to protect bitstreams.
However, prior work has shown that these protections can often be broken or circumvented by (i)~side channel attacks based on leaked information through power or electromagnetic radiation~\cite{moradi2011vulnerability, swierczynski2015physical, DBLP:journals/tches/HettwerLFGG21}, (ii)~optical attacks leaking keys~\cite{DBLP:conf/ccs/TajikLSB17, DBLP:journals/tches/LohrkeTKBS18}, or (iii)~attacks exploiting bugs in the implementation of security mechanisms~\cite{DBLP:conf/uss/Ender0P20, DBLP:conf/fccm/EnderLMP22, DBLP:journals/tches/EnderHFMP24}.
We therefore consider protection mechanisms out of scope and, like most prior work, focus on unencrypted bitstreams.

\paragraph{\blackbox{F4} Bitstream Conversion.}
In the literature, converting a bitstream into a gate-level netlist comprises two steps: (i)~recovering logic elements and (ii)~reconstructing routing.
The behavior of logic elements can be inferred from the corresponding configuration bits in the reversed bit mapping.
This includes reconstructing Boolean functions from truth tables, either through Boolean minimization~\cite{yu2021multi, jeong2018extract} or machine learning~\cite{narwariya2024revbit}.

Interconnections of logic elements are implemented using sequences of routing segments, which can be controlled via the bitstream at physical junctions using programmable switches.
To reconstruct such connections, the individual segments must be merged into coherent nets, \eg, through a \ac{BFS}~\cite{zhang2019comprehensive}.
Some elements have ambiguous configurations.
For example, distinguishing between an unused component and a component in default configuration may be impossible~\cite{ding2013deriving}.
In such cases, surrounding elements must be taken into consideration~\cite{zhang2019comprehensive}.

A related branch of research has successfully inferred high-level information about the design, such as the presence of specific \ac{IP} cores or functional blocks, by analyzing the bitstream directly without performing netlist recovery.
Initial work identifies \ac{IP} cores based on \acp{LUT} contents~\cite{ziener2006identifying}, while more recent research uses machine-learning techniques~\cite{mahmood2019ip}, or operates on bitstreams transformed into images~\cite{chen2021deep}.

\subsubsection{Technical Challenges in \texorpdfstring{\acs{FPGA}}{FPGA} Reverse Engineering and Ways Forward}

\paragraph{(Verified) Completeness.}
Modern \acp{FPGA} are highly complex and include device-specific features beyond the classical configurable elements, \eg, hard \ac{IP} such as \ac{DSP} blocks and PCIe interfaces.
A major technical challenge is achieving and verifying the completeness of the recovered bitstream format in a black-box setting.
Solving this requires additional information, such as power-up default values, and a comprehensive understanding of the configuration process~\cite{williams2023determining}.
Although some studies claim perfect recovery of the entire bitstream format, they are only validated on minimal benchmarks (see~\autoref{table::fpga_re::results}).

\proposalbox{
Evaluate on a set of bitstreams of varying complexity covering different design primitives and device settings.
Define a suitable coverage metric that captures the proportion of the bitstream that was successfully reverse-engineered, accounting for the complexity and semantic relevance of different bitstream components.
}

\paragraph{Reliance on Deprecated Tooling and Narrow Focus.}
Many existing approaches rely on the ability to control each configurable element in isolation~\cite{zhang2019comprehensive, choi2020reverse, zhang2023bitfree}.
While older \ac{EDA} tools such as AMD\footnote{We use AMD to refer to devices formerly branded as Xilinx, consistent with their current ownership under AMD.} ISE, which has been deprecated since 2013, often allowed for such fine-grained control, it is no longer supported in modern tools such as AMD Vivado~\cite{zhang2023bitfree}.
Although some progress has been made in adapting legacy reverse-engineering workflows to more recent toolchains, the underlying lack of fine-grained control remains unresolved and continues to hinder reverse engineering, particularly with respect to routing~\cite{choi2024approaches}.
Consequently, many proposed methods cannot be applied to current-generation devices, including the speed-ups proposed by Zhang~\etal~\cite{zhang2023bitfree}.
Furthermore, existing research on bitstream reverse engineering exhibits a disproportionate focus on AMD \acp{FPGA}, see~\autoref{table::fpga_re::results}.
The resulting over-representation of a single vendor could stem from the lack of public tooling and low-level documentation from other vendors.
The absence of other manufacturers and the lack of generalization leave a research gap.

\proposalbox{
Focus on modern \ac{FPGA} architectures and toolchains, especially from Intel, which remain drastically underrepresented relative to their market share.
Since modern toolchains prevent fine-grained isolation of individual configurable elements, straightforward one-to-one mappings became infeasible.
Therefore, previous state-of-the-art correlation-based approaches present a tool-agnostic path forward.
}

\paragraph{Scattered Information.}
\ac{FPGA} bitstream documentation is used not only for reverse engineering but also for design applications.
Therefore, relevant work is scattered across communities---hardware reverse engineers, open-source designers~\cite{10.1007/11549703_6, 9415542, prjxray, icestorm, trellis}, groups specializing on bitstream manipulation~\cite{dangpham2017bitman, manev2022byteman, kalte2005replica, kalte2006replica2pro}, and researchers investigating \ac{FPGA} bitstream security~\cite{swierczynski2015physical, moradi2011vulnerability, DBLP:journals/tches/EnderHFMP24}---leading to similar approaches and tools being developed in parallel.

\proposalbox{
Facilitate exchange among the hardware reverse engineering, open-source design, and bitstream manipulation communities, and adapt their existing solutions.
}

\paragraph{Limitations in Methodology and Validation.}
Overall, research on \ac{FPGA} reverse engineering is of mixed quality.
Often, works omit crucial details, rendering independent reproduction or use of their techniques challenging, even for domain experts, further worsened by the fact that only two publications release reference implementations.
Evaluations often validate only a subset of the stated claims, limiting confidence in the reported results and obscuring the rigor of the underlying methods.
Moreover, some publications offer limited novelty by revisiting problems already solved in prior work.
These issues impede cumulative progress and may explain the lack of \ac{FPGA} reverse engineering publications at top-tier venues.
This scarcity may also be supported by vendors’ efforts to protect their \ac{IP} through licensing restrictions that try to legally prohibit bitstream analysis.
Collectively, these limitations may explain the lack of publicly available, peer-reviewed academic tools capable of recovering arbitrary netlists from \ac{FPGA} bitstreams.
For now, open-source initiatives such as Project X-Ray~\cite{prjxray} and IceStorm~\cite{icestorm}, while not achieving \SI{100}{\percent} completeness, represent the only available and usable solutions to this end.

\proposalbox{
Put forth good academic practices as authors and enforce them rigorously as reviewers, especially in regard to reproducibility, methodological detail, and validation.
}

\subsection{Netlist Reverse Engineering}

Both \blackbox{I5} Netlist Extraction and \blackbox{F4} Bitstream Conversion yield a gate-level netlist consisting of combinational gates (\eg, AND, OR, XOR), sequential elements (\eg, \acp{FF} and latches), and their interconnections.
Such a netlist is often described as a \enquote{sea of gates}~\cite{albartus2020dana}, reflecting its scale and structural complexity.
Netlist reverse engineering aims to interpret this sea of gates by recovering higher-level information and reintroducing abstractions removed during design and manufacturing.
Its goals range from identifying regions of interest (\eg, locating fault-injection targets)~\cite{leander2024hawkeye} to understanding implemented algorithms~\cite{klix2024stealing} (\eg, verifying the absence of backdoors in third-party designs)~\cite{meade2018old}.
We refer to Xu~\etal~\cite{cryptoeprint:2026/531} for a complementary comparison of netlist reverse engineering algorithms.
Here, we instead focus on the sequence of reverse engineering steps and the dependencies between them.

\subsubsection{Techniques for Netlist Reverse Engineering}
\label{hre_sok::section::netlist_rq1}

\begin{figure}[htb]
    \centering
    \includegraphics[width=0.7\linewidth]{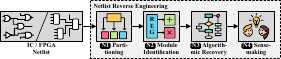}
    \caption{Netlist reverse engineering flow.}
    \label{hre_sok::figure::netlist_flow}
\end{figure}

Netlist reverse engineering can be broadly organized into four steps (see \autoref{hre_sok::figure::netlist_flow}): \blackbox{N1}~Partitioning, \blackbox{N2}~Module Identification, \blackbox{N3}~Algorithmic Recovery, and \blackbox{N4}~Sensemaking.

\paragraph{\blackbox{N1} Partitioning.}
Commonly, the netlist is partitioned by assigning gates to groups that correspond to functional cores, structures such as registers and arithmetic subcircuits, or just categories such as datapath and control path~\cite{subramanyan2014reverse}.

\textit{Cores.} Several works suggest that connectivity is denser within a module than across module boundaries.
Based on this assumption, Cakir~\etal cluster gates using eigenvalue decompositions of the adjacency matrix~\cite{cakir2018reverse, cakir2019revealing}, Weber~\etal apply graph clustering algorithms to gate-level netlists~\cite{weber2022analysis}, and Couch~\etal use an approximation of the NCut criterion~\cite{couch2016functional}.
Werner~\etal employ graph clustering, refined with structural properties, and combined with visual netlist graph inspection~\cite{werner2018reverse}.
Another cue for separating cores is physical proximity: due to routing and timing constraints, related cells are often placed nearby.
Such placement information can be used as input to a \ac{GNN} that partitions the netlist into clusters~\cite{hong2023graphclusnet}.

\textit{Word-Level Structures.} Partitioning can also focus on smaller structures, grouping gates of the same word-level register or (arithmetic) operation.
Physical proximity helps cluster single-bit elements into words~\cite{nathamuni-venkatesan2023wordlevel}.
Many methods assume that each bit of a datapath word undergoes the same computation, so their surrounding logic should be structurally similar.
Similarity can be measured by comparing fan-in cuts~\cite{meade2018old, tashjian2015using}, computing graph edit distances over local neighborhoods~\cite{nathamuni-venkatesan2023wordlevel}, or using BERT-style models~\cite{zhang2025rebert}.
To better separate words with similar fan-in structure, Meade~\etal incorporate additional structural features and apply \ac{PCA} to embed bits of the same word nearby in a high-dimensional space~\cite{meade2018old}.

A complementary heuristic groups gates by shared control signals: since those steered by the same control typically belong to the same word~\cite{chisholm1999understanding, albartus2020dana, tashjian2015using, fyrbiak2020graph}.
Albartus~\etal extend this approach by potentially incorporating expected register sizes and examining shared predecessor and successor groups, based on the insight that all bits of a word-level register typically connect to the same surrounding registers and belong to the same register stage~\cite{albartus2020dana}.
Subsequent work refines this by using known word groupings from other analyses~\cite{klix2024stealing}, and leveraging \ac{FPGA} primitives that already expose word-level interfaces~\cite{mckendrick2022leveraging}.
The shared successor and predecessor assumption can also be used to propagate known word boundaries, \eg, by symbolically evaluating the combinational logic around an already identified word~\cite{li2013wordrev, subramanyan2013reverse, subramanyan2014reverse}.

\textit{Control Logic.} Many approaches partition circuits into datapath and control path by identifying \ac{FSM} \acp{FF} that belong to the control path.
Here, the datapath-similarity assumption is inverted: \acp{FF} whose fan-in structure is dissimilar to word-level datapath logic are likely control logic.
This can be done by structural comparison of fan-in cuts~\cite{meade2016gatelevel, brunner2019improving} or by comparing the same cuts at a functional level~\cite{geist2020relicfun}.
Another heuristic assumes control logic connects to many parts of the circuit.
Chowdhury~\etal use this assumption and calculate graph centrality metrics as inputs to a \ac{GNN} that classifies \acp{FF} as control or datapath~\cite{chowdhury2021reignn}.
A specialized approach targeting block ciphers analyzes the circuit under induced \ac{FF} faults, assuming faults in control logic produce different effects than those in datapath registers~\cite{mildner2024faultsimulationbased}.

\paragraph{\blackbox{N2} Module Identification.}
Next, functionality is assigned either to modules obtained from partitioning or through an interplay of partitioning and module identification.

\textit{Datapath.} Most commonly, datapath functionality is identified by exploiting structural similarities between similar high-level modules within the implemented circuit.
This enables the creation of libraries comprising known templates, which are then matched against unknown (parts of) netlists.
The central challenge is thus measuring similarity between candidate subcircuits and reference implementations.

Early work explored subgraph isomorphism on subcircuits or whole designs~\cite{chisholm1999understanding, bouchaour2011method}, but its computational cost motivates alternative comparisons.
These include handcrafted subgraph embeddings combined with classical machine learning~\cite{baehr2019machine}, graph-similarity measures~\cite{fyrbiak2020graph} and embeddings~\cite{couch2016functional}, and hyper-dimensional graph encoding~\cite{genssler2024circuithd}.
Other approaches compute node embeddings from local neighborhoods and sample them for machine-learning-based classification~\cite{fayyazi2019deep} or match nodes individually against a pattern library~\cite{baehr2019machine, skipper2020recon, dahl2023iprec}.
Some techniques use \acp{GNN} to learn node embeddings and predict the operation a node is part of, trained on pattern libraries~\cite{alrahis2022gnnre, bucher2022appgnn, pula2023relutgnn, zhao2025adversarial, he2021graph, wang2023efficient}.
Beyond structural matching, extracted cuts can be compared functionally to find semantic matches, independent of structural implementation~\cite{shi2012extracting, li2013wordrev, subramanyan2013reverse, subramanyan2014reverse}.

Not all methods rely on generic matching.
Several approaches search for domain-specific characteristics, such as \ac{FF} structures for \acp{LFSR} or counters~\cite{subramanyan2013reverse, subramanyan2014reverse, wallat2017look}, or carry chains hinting at arithmetic operations~\cite{klix2024stealing, narayanan2023reverse}.
On architectures without dedicated carry primitives, half adders can be reconstructed from XOR patterns~\cite{stoffel2004equivalence, wei2015universal}.
Related techniques reintroduce (half-)adder primitives to ease equivalence checking, using functional bitslice analysis~\cite{mahzoon2022revsca20}, rewriting via e-graphs~\cite{yin2025boole}, or \acp{GNN}~\cite{wu2023gamora}.
Similarly, Leander~\etal exploit block cipher structure to detect cryptographic implementations~\cite{leander2024hawkeye}.

Another issue is reconstructing the order of subcircuit \ac{IO} signals and mapping them to operands or templates.
For arithmetic subcircuits, this \ac{IO} mapping can be deduced from structural or functional patterns.
A common cue is that the least significant input bits affect many output bits, whereas the most significant bits influence fewer outputs~\cite{ho2023wolfex, he2021graph, wang2023efficient, klix2024stealing, gascon2014templatebased}.
Other approaches evaluate the circuit on all-zero or one-hot/cold inputs~\cite{doom1998identifying, chisholm1999understanding, shi2012extracting}.
Some techniques even use such behavioral signatures to perform module identification and \ac{IO} mapping simultaneously~\cite{li2012reverse, soeken2015simulation}.

Given a functional hypothesis for an unknown subcircuit, formal verification can be used to confirm equivalence.
This includes \ac{SAT} and \ac{SMT} solving~\cite{yu2019formal, klix2024stealing, wei2015universal, delozier2024thirdparty, gascon2014templatebased}, \ac{QBF} solving~\cite{li2013wordrev, subramanyan2014reverse, gascon2014templatebased, narayanan2023reverse}, \ac{BDD} matching~\cite{doom1998identifying, chisholm1999understanding}, \ac{SCA}-based techniques~\cite{mahzoon2019revsca, mahzoon2022revsca20, yin2025boole}, polynomial rewriting~\cite{ho2023wolfex, yu2016automatic}, linear coefficient fitting~\cite{ho2023wolfex}, and symbolic regression~\cite{ho2023wolfex}.

\textit{Control Logic.} Existing work also aims to identify \acp{FF} that form an \ac{FSM} state register.
Such methods often exploit the interconnections of state bits: since the next state is computed based on the current one, state registers often form cyclic dependencies.
Accordingly, such approaches search for \acp{SCC} or feedback paths~\cite{chisholm1999understanding, kibria2022fsmx, kibria2023fsmxultra, shi2010highly, brunner2022humanreadable, wallat2019highway, portillo2019rertl}.
State registers can also be identified by analyzing whether \ac{FF} fan-outs connect to control nets such as enable, reset, or MUX select~\cite{klix2024stealing, muthukumaran2023reverse, shi2010highly}.

\paragraph{\blackbox{N3} Algorithmic Recovery.}

A netlist lifted to higher-level modules can be described as a \enquote{sea of modules}, reflecting the added abstraction and reduced complexity~\cite{subramanyan2014reverse}.
However, to later understand the implemented computations, the challenge is to recover an algorithmic description from these modules.
For algorithmic recovery, the reverse engineer is often faced with data-dependent sequential behavior.
They need to combine recovered control-path and datapath information to compile a more explicit representation.

\textit{\ac{FSM} State Graph.} For control logic, \ac{FSM} state graphs are reconstructed from the \ac{FSM} subcircuit by analyzing state transitions to produce a state traversal graph.
This is done by brute-force enumeration~\cite{wallat2019highway} or by analyzing next-state logic via \ac{ATPG}~\cite{kibria2023fsmxultra}, \ac{ROBDD}, or \ac{SAT}-based reasoning~\cite{portillo2019rertl, meade2017ip, meade2019neta, meade2016netlist}.

\textit{Control and Datapath Combination.} Integrating results from multiple reverse-engineering steps remains difficult and is rarely addressed explicitly.
Narayanan~\etal adapt existing techniques and combine them with novel methods to extract an \ac{RTL} description~\cite{narayanan2023reverse}.
Klix~\etal present a solution toward achieving a higher-level algorithm description comprising two components: (i)~manual tracing of simulation results obtained from real-world inputs, and (ii)~an automated hybrid of concrete and symbolic simulation that outputs a Python replication of the circuit behavior~\cite{klix2024stealing}.

\paragraph{\blackbox{N4} Sensemaking.}
Sensemaking is the final step after successive abstraction recovery: interpreting the recovered algorithm in its operational context.
This stage is largely unexplored; so far, prior work has only proposed manual analysis by domain experts~\cite{klix2024stealing}.

\subsubsection{Technical Challenges in Netlist Reverse Engineering and Ways Forward}

\paragraph{Hurdles for End-to-End Automation.}
Manual netlist reverse engineering is highly time-consuming, making automation essential for practical and economic viability~\cite{fyrbiak2020graph,leander2024hawkeye}.
While many recent works propose algorithms for individual subproblems, end-to-end automation remains elusive.
Isolated solutions are often difficult to combine: they may be non-interoperable, rely on incompatible assumptions (\eg, specific architectures or design layouts), or require idealized input from preceding steps.
Furthermore, many approaches depend on hand-tuned parameters or thresholds, preventing full automation~\cite{azriel2021survey, nathamuni-venkatesan2023wordlevel, chowdhury2021reignn,meade2018old, brunner2019improving, kibria2023fsmxultra}.
Effective integration requires standardized interfaces and data formats for exchanging intermediate results and a common framework.
While such frameworks exist~\cite{wallat2019highway, skouson2020netlist}, adoption is limited.
Even with a unified toolchain, human experts must interpret and validate intermediate results~\cite{chisholm1999understanding,subramanyan2014reverse,klix2024stealing}, highlighting the need for tools that integrate analyst expertise and for further study of how analysts perform such tasks~\cite{fyrbiak2017hardware,DBLP:journals/tochi/WiesenBWPR23}.

\proposalbox{
Enable and evaluate the interplay between existing algorithms, and design novel approaches with integration into a semi-automated pipeline in mind.
To achieve this, design standardized APIs, adopt shared frameworks, avoid hand-tuned parameters, and develop efficient human-in-the-loop interaction.
}

\paragraph{Focus on Early Stages.}
Existing work often targets early-stage tasks, particularly partitioning and module identification~\cite{azriel2021survey}.
These stages decompose a netlist into (identified) word-level modules and possibly even larger functional blocks.
However, many important applications, such as vulnerability discovery, require insights beyond the word level.
A holistic design understanding is only achieved through algorithmic recovery and higher-level sensemaking.
This, in turn, requires jointly interpreting datapath and control logic through additional abstractions, yet these later stages remain largely unexplored.

\proposalbox{
Develop methods that raise abstraction beyond word-level netlists or \ac{RTL}, to achieve a specification-level understanding.
Promising directions include dynamic analysis and joint interpretation of recovered datapath modules and \ac{FSM} state graphs.
}

\paragraph{Overly Idealized Assumptions.}
Many proposed algorithms rely on assumptions that rarely hold in practice.
Almost all assume error-free netlists~\cite{chisholm1999understanding,werner2018reverse, baehr2019machine,fyrbiak2020graph, hong2024mlconnect}, which is unrealistic given the error-prone netlist recovery process for \acp{IC}.
Some module identification approaches further assume perfect partitioning into (yet unidentified) modules~\cite{subramanyan2014reverse} or a known \ac{IO} mapping from library candidates to such modules~\cite{hansen1999unveiling,chisholm1999understanding,gascon2014templatebased,keshavarz2018survey}, both challenging and error-prone tasks on their own.
Other algorithms are evaluated on unoptimized netlists---\eg, without logic merging across module boundaries~\cite{alrahis2022gnnre}---or on task-specific netlists that only include arithmetic components, sidestepping cases that break their assumptions and risking false detections or excessive runtimes~\cite{ho2023wolfex, liu2025widegate}.
Several control-logic identification methods assume \acp{FSM} form a small, separable \ac{SCC}, yet real designs often collapse control and datapath into a single large \ac{SCC} due to feedback loops~\cite{klix2024stealing}.
Together, these assumptions make it difficult to assess algorithm effectiveness under realistic conditions.

\proposalbox{
Evaluate novel and existing techniques on complex and erroneous netlists that were synthesized using different optimization strategies.
Realistic errors could be synthetically generated based on common issues observed for \ac{IC} and \ac{FPGA} netlist recovery.
A unified, sufficiently challenging benchmark suite exposes algorithms with overly idealized assumptions by poor generalization on realistic designs.
}

\section{Evaluation of \texorpdfstring{\acs{HRE}}{HRE} Artifacts}
\label{hre_sok::section::artifact_eval}

Across all three subdomains, challenges identified in \autoref{section:results} point to a common pattern: \ac{HRE} progress is constrained by isolated research and limited reuse of tools and datasets, reflecting a limited culture of building upon prior work. 
To confirm this, we examine research artifacts, whose availability and reuse are essential to such a culture.

Out of the 187 papers, 31 (\SI{17}{\percent}) report publicly available artifacts.
As two papers reference the same artifact, this corresponds to 30 unique artifacts.
Despite a recent increase in artifact publication, as illustrated in \autoref{fig:intro:pub-artifact-dist}---\SI{24}{\percent} for papers published between 2021 and 2025 compared to \SI{9}{\percent} between 1986 and 2020---the integration of artifacts into \ac{HRE} research remains limited.
We summarize our findings in \autoref{table:artifact}.

\begin{table}[htb]
    \caption{
Detailed per-artifact evaluation results for all publicly available \ac{HRE} artifacts identified in our corpus. 
Artifacts are grouped by domain and evaluated along the dimensions of \textit{Availability}, \textit{Functionality} (documentation, completeness, exercisability), and \textit{Reproducibility}. 
Symbol encodings represent graded assessment outcomes (see legend).}
    \label{table:artifact}
    \medskip
    \centering
    \scriptsize
    \setlength{\tabcolsep}{4pt}

    \begin{minipage}[t]{0.40\textwidth}
        \vspace{0pt} 
        \centering
        \begin{tabular}{l ccc ccc c}
            \toprule
            \multirow[t]{2}{*}{\textbf{}} &
            \multicolumn{3}{c}{\textbf{Avail.}} &
            \multicolumn{3}{c}{\textbf{Funct.}} &
            \multirow[t]{2}{*}{\textbf{Reprod.}} \\
            \cmidrule(lr){2-4} \cmidrule(lr){5-7} \cmidrule(lr){8-8}
            \addlinespace[0.3em]
            & Tool & Image & Other & D & C & E & \\
            \midrule
            \multicolumn{8}{l}{\textsc{IC Reverse Engineering Artifacts}}\\
            \midrule
            \rowcolor{gray!20}\cite{rajarathnam2020regds} & \CIRCLETHREEQUARTER & \textemdash & \CIRCLETHREEQUARTER & \CIRCLETHREEQUARTER & \CIRCLEFULL & \CIRCLETHREEQUARTER & \ding{55} \\
            \cite{wilson2021refics} \\ \cite{wilson2022refics} & \CIRCLETHREEQUARTER & \CIRCLETHREEQUARTER & \CIRCLETHREEQUARTER & \CIRCLEHALF & \CIRCLEFULL & \CIRCLEHALF & \ding{55} \\
            \rowcolor{gray!20}\cite{burian2022automated} & \textemdash & \CIRCLEFULL & \textemdash & \CIRCLEFULL & \CIRCLEFULL & \textemdash & \ding{55} \\
            \cite{wilson2023secure} & \CIRCLETHREEQUARTER & \textemdash & \textemdash & \CIRCLEEMPTY & \CIRCLEFULL & \CIRCLETHREEQUARTER & \ding{55} \\
            \rowcolor{gray!20}\cite{liu2024novel} & \CIRCLEEMPTY & \CIRCLEEMPTY & \CIRCLEEMPTY & \textemdash & \textemdash & \textemdash & \textemdash \\
            \cite{marazzi2024hifidram} & \textemdash & \CIRCLETHREEQUARTER & \CIRCLETHREEQUARTER & \CIRCLEFULL & \CIRCLEFULL & \textemdash & \textemdash \\
            \rowcolor{gray!20}\cite{xiao2024denoising} & \CIRCLEEMPTY & \CIRCLEEMPTY & \textemdash & \textemdash & \textemdash & \textemdash & \textemdash  \\
            \cite{xiao2024tadensenet} & \textemdash & \CIRCLETHREEQUARTER\;+\;\CIRCLEEMPTY & \textemdash & \CIRCLEEMPTY & \CIRCLEHALF & \textemdash & \ding{55} \\
            \rowcolor{gray!20}\cite{zhu2025genetic} & \CIRCLETHREEQUARTER & \CIRCLETHREEQUARTER & \textemdash & \CIRCLETHREEQUARTER & \CIRCLEFULL & \CIRCLEFULL & \ding{55} \\
            \cite{vanderlinden2025simple} & \CIRCLETHREEQUARTER & \textemdash & \textemdash & \CIRCLETHREEQUARTER & \CIRCLEFULL & \CIRCLEHALF & \textemdash \\
            \rowcolor{gray!20}\cite{cheng2025unsupervised} & \textemdash & \CIRCLEHALF & \textemdash & \CIRCLEFULL & \CIRCLEFULL & \textemdash & \ding{55} \\
            \cite{rothaug2025advancing} & \CIRCLETHREEQUARTER & \CIRCLEFULL & \CIRCLETHREEQUARTER & \CIRCLETHREEQUARTER & \CIRCLEFULL & \CIRCLEFULL & \ding{51} \\
            \midrule
            \multicolumn{8}{l}{\textsc{FPGA Reverse Engineering Artifacts}}\\
            \midrule
            \rowcolor{gray!20}\cite{benz2012bil} & \CIRCLETHREEQUARTER & \textemdash & \textemdash & \CIRCLEFULL & \CIRCLEFULL & \CIRCLETHREEQUARTER & \ding{51} \\
            \cite{kashani2022bitfiltrator} & \CIRCLETHREEQUARTER & \textemdash & \CIRCLETHREEQUARTER & \CIRCLEFULL & \CIRCLEFULL & \CIRCLEFULL & \textemdash \\     
            \midrule
            \multicolumn{8}{l}{\textsc{Netlist Reverse Engineering Artifacts}}\\
            \midrule
            \rowcolor{gray!20}\cite{hansen1999unveiling} & \textemdash & \textemdash & \CIRCLEHALF & \CIRCLEFULL & \CIRCLEFULL & \textemdash & \textemdash \\
            \cite{gascon2014templatebased} & \CIRCLEEMPTY & \textemdash & \CIRCLEEMPTY & \textemdash & \textemdash & \textemdash & \textemdash\\
            \rowcolor{gray!20}\cite{soeken2015simulation} & \CIRCLEEMPTY & \textemdash & \CIRCLEEMPTY & \textemdash & \textemdash & \textemdash & \textemdash \\
            \cite{meade2019neta} & \CIRCLETHREEQUARTER & \textemdash & \textemdash & \CIRCLETHREEQUARTER & \CIRCLEFULL & \CIRCLEFULL & \textemdash \\
            \rowcolor{gray!20}\cite{wallat2019highway} & \CIRCLETHREEQUARTER & \textemdash & \textemdash & \CIRCLEFULL & \CIRCLEFULL & \CIRCLEFULL & \textemdash \\
            \cite{skouson2020netlist} & \CIRCLETHREEQUARTER & \textemdash & \textemdash & \CIRCLETHREEQUARTER & \CIRCLEFULL & \CIRCLEFULL & \textemdash \\
            \rowcolor{gray!20}\cite{albartus2020dana} & \CIRCLETHREEQUARTER & \textemdash & \CIRCLETHREEQUARTER & \CIRCLETHREEQUARTER & \CIRCLEFULL & \CIRCLETHREEQUARTER & \ding{55} \\
            \cite{mahzoon2022revsca20} & \CIRCLETHREEQUARTER & \textemdash & \CIRCLETHREEQUARTER & \CIRCLETHREEQUARTER & \CIRCLEFULL & \CIRCLETHREEQUARTER & \ding{51}\\
            \rowcolor{gray!20}\cite{alrahis2022gnnre} & \CIRCLETHREEQUARTER & \textemdash & \CIRCLETHREEQUARTER & \CIRCLETHREEQUARTER & \CIRCLEFULL & \CIRCLEHALF & \ding{55} \\
            \cite{bucher2022appgnn} & \CIRCLETHREEQUARTER & \textemdash & \CIRCLETHREEQUARTER & \CIRCLETHREEQUARTER & \CIRCLEFULL & \CIRCLEHALF & \ding{55} \\
            \rowcolor{gray!20}\cite{sisco2023loop} & \CIRCLEFULL & \textemdash & \CIRCLEFULL & \CIRCLEHALF & \CIRCLEFULL & \CIRCLETHREEQUARTER & \ding{51} \\
            \cite{wu2023gamora} & \CIRCLETHREEQUARTER & \textemdash & \CIRCLETHREEQUARTER & \CIRCLEHALF & \CIRCLEFULL & \CIRCLEHALF & \ding{51} \\
            \rowcolor{gray!20}\cite{hong2023graphclusnet} & \CIRCLETHREEQUARTER & \textemdash & \CIRCLETHREEQUARTER & \CIRCLEHALF & \CIRCLEFULL & \CIRCLETHREEQUARTER & \ding{55} \\
            \cite{klix2024stealing} & \CIRCLEFULL & \textemdash & \CIRCLEFULL & \CIRCLETHREEQUARTER & \CIRCLEFULL & \CIRCLETHREEQUARTER & \ding{51} \\
            \rowcolor{gray!20}\cite{leander2024hawkeye} & \CIRCLEFULL & \textemdash & \CIRCLEFULL & \CIRCLETHREEQUARTER & \CIRCLEFULL & \CIRCLETHREEQUARTER & \ding{51} \\
            \cite{liu2025widegate} & \CIRCLEEMPTY & \textemdash & \textemdash & \textemdash & \textemdash & \textemdash & \textemdash \\
            \bottomrule
        \end{tabular}
    \end{minipage}
    \hfill
    \begin{minipage}[t]{0.39\textwidth}
        \vspace{0pt}
        \raggedright
        \begin{description}[style=nextline, leftmargin=0pt]
            \item[\underline{Availability}]
                {\CIRCLEFULL}{ In permanent storage}
                \\{\CIRCLETHREEQUARTER}{ In non-permanent storage (\eg, personal website, GitHub, \dots)}
                \\{\CIRCLEHALF}{ Upon request or via online search}
                \\{\CIRCLEEMPTY}{ Unavailable or no author response}
                \\{\textemdash}{ Not contained in artifact}
            
            \item[\underline{Functionality}]
                \textbf{D}ocmentation
                \\{\CIRCLEFULL}{ Clear documentation}
                \\{\CIRCLETHREEQUARTER}{ Minor issues }
                \\{\CIRCLEHALF}{ Major issues}
                \\{\CIRCLEEMPTY}{ Not usable}
                \\{\textemdash}{ Artifact not available}

                 \textbf{C}ompleteness
                \\{\CIRCLEFULL}{ Matches paper description}
                \\{\CIRCLETHREEQUARTER}{ Minor differences from paper}
                \\{\CIRCLEHALF}{ Major differences from paper}
                \\{\CIRCLEEMPTY}{ Does not match paper description}
                \\{\textemdash}{ Artifact not available}

                \textbf{E}xercisability
                \\{\CIRCLEFULL}{ Out of the box}
                \\{\CIRCLETHREEQUARTER}{ After minor effort}
                \\{\CIRCLEHALF}{ After major effort or partially exercisable}
                \\{\CIRCLEEMPTY}{ Not exercisable despite major effort}
                \\{\textemdash}{ Artifact not available or contains no tool}
            
            \item[\underline{Reproducibility}]
                {\ding{51}}{ Some results successfully reproduced}
                \\{\ding{55}}{ No results could be reproduced}
                \\{\textemdash}{ Artifact not available or not exercisable or no results to be reproduced}
        \end{description}
    \end{minipage}
\end{table}

\paragraph{Availability.} 

Of the 30 artifacts identified in our corpus, 24 were fully available, one was partially available, and five were inaccessible.
Permanent storage appeared only 5 times, first in 2022, but its adoption is encouraging given that three non-permanently hosted artifacts are already inaccessible.
While 23 artifacts used a single storage method---permanent storage, non-permanent storage, or access upon request---two artifacts were distributed across multiple storage types.
One of these relied on non-permanent hosting and access upon request, but our request was never answered.

\paragraph{Functionality.}
Across the 25 available artifacts, 2 provided no usable guidance (containing no or non-applicable instructions), and 4 exhibited major \textit{documentation} gaps (\eg, hard-coded absolute paths, no execution instructions, or missing dependency lists).
The remaining 19 artifacts described a workable \enquote{path to first success,} with only minor friction such as adding a missing package, a build flag, or adjusting a configuration or file path.
In terms of \textit{completeness}, all artifacts were fully delivered, except for one, where part of the artifact had to be requested, but was not provided.
Regarding \textit{exercisability}, our evaluation of the 20 available tools showed that all were executable, but only 6 could be executed without issues.
For 14 artifacts, execution typically required adjustments to the setup---often due to incomplete or unclear documentation---resulting in a trial-and-error approach.
All encountered issues, along with resolution strategies, are documented for each artifact in our supplementary material. 
Given our best-effort evaluation without author support, these results are a reasonable lower bound of the artifacts' real-world usability.

\paragraph{Reproducibility.} 
Of 18 papers with available artifacts, testable results, and executable tools, we could reproduce key results of seven papers.
The remaining attempts failed primarily due to missing evaluation materials (\eg, required binaries, netlists, bitstreams, \ac{IC} images, or labeled datasets) or because the 
required evaluation tools were unavailable
(\eg, for computing the error rate between segmented images and labeled data, or for calculating evaluation scores).
While we acknowledge that not all papers in our corpus contain verifiable claims, we consider 
seven reproducible papers (\SI{4}{\percent}) out of 187 to be exceptionally low.

\section{Cross-Cutting Opportunities for Scalable \texorpdfstring{\acs{HRE}}{HRE} Research}
\label{hre_sok::section::rq3_overcoming}

Across all three subdomains, our analysis shows progress is not only limited by isolated technical challenges but even more by three structural barriers that cut across \ac{IC}, \ac{FPGA}, and netlist reverse engineering: scarce reusable artifacts, which undermine reproducibility; missing benchmarks, ground truth, and evaluation standards, which hinder comparison; and limited legal clearance and clarity, which restricts data sharing and collaboration.

These barriers recur because
\ac{HRE} is shaped by a common stakeholder landscape---academia, government institutions, defense-affiliated organizations, and specialized sectors of the semiconductor industry---whose interests span the full reverse-engineering workflow. 
In this setting, intellectual-property and national-security concerns can discourage open-science practices, a pattern which, for example, manifests in the low artifact-publication rates in our corpus: among papers with defense-related funding or affiliation, artifact publication rates fall to 10\%, and among those with industry affiliations to 7\%, compared to 19\% for all remaining papers. 
Yet these same stakeholders stand to gain the most from addressing structural barriers.
A government agency cannot independently verify contractor results---because tooling and data are proprietary---and therefore must ultimately accept their claims based on trust.
This is precisely the verification gap that public artifacts and shared benchmarks are designed to close.
Similarly, industry and government benefit from more trained practitioners and validated methods, while academia can shift from repeatedly rebuilding missing foundations to producing cumulative knowledge.
Addressing the three core structural issues is therefore not only academically desirable but aligned with the long-term interests of actors who fund, conduct, and rely on \ac{HRE} research. 

Because the same barriers recur across the whole field, this section discusses three cross-cutting \emph{opportunities} and derives stakeholder-specific recommendations for each.
These opportunities are not fully independent; legal clearance and clarity (Opportunity~III) are among the preconditions for both reproducibility (Opportunity~I) and comparability (Opportunity~II), since artifact sharing and benchmark publication require addressing \ac{IP} and export-control matters.
We nonetheless treat them separately,  as each calls for distinct actions and involves different stakeholder trade-offs.
We tag proposals by the effort their adoption requires:
\termbox{ST}~short-term steps that individual researchers or venues can adopt within current structures; \termbox{MT}~mid-term changes requiring institutional shifts in incentives or funding; and \termbox{LT}~long-term goals contingent on broader policy or market change.

\paragraph{Opportunity I: Value Reproducibility and Reusability.}
Many contributions remain difficult to reproduce and build upon: code is often unpublished or fragile, \ac{HRE} pipelines are bespoke, and as a result, technical progress remains isolated.
We attribute this to structural factors, including missing academic incentives, limited awareness, insufficient industry cooperation, and legal constraints on publication.
This situation is particularly harmful in \ac{HRE}, where intermediate results naturally feed later pipeline stages.
\ac{HRE} lacks widely adopted, publicly available frameworks for integrating new methods, comparable to the rich open-source ecosystem in \ac{SRE}, with tools like Ghidra.
Promising early efforts, such as Degate~\cite{degate}, have largely stagnated.
Meanwhile, \ac{FPGA}-focused tools like Project X-Ray~\cite{prjxray} remain limited to individual device families.
For netlist reverse engineering, HAL~\cite{fyrbiak2019hal} represents a notable exception and has gained some traction~\cite{DBLP:journals/corr/abs-2512-14139}, yet it still lacks widespread adoption.

\stakeholderProposals{Venues and Academic Incentives}{
Improving reproducibility and reuse requires aligning academic incentives with artifact-centric research practices.
\termbox{ST}~While some conferences have demonstrated effective mechanisms through mandatory artifact evaluation, artifact publication and long-term archival should become the default for all \ac{HRE}-relevant venues.
\termbox{ST}~The community should also explicitly reward engineering work through tool and infrastructure tracks and through mechanisms that promote long-term maintenance, \eg, recognition for impactful, living artifacts~\cite{acsac2025artifacts}.
\termbox{MT}~These priorities should extend beyond venues and be reflected in Ph.D. program design, hiring, and funding decisions, treating tools and curated datasets as first-class research outputs.
}

\stakeholderProposals{Individual Researchers}{
\termbox{ST}~Within such incentive-aligned environments, researchers should select venues that facilitate rigorous artifact evaluation practices.
\termbox{ST}~For long-term reproducibility, researchers should package artifacts in containers or virtual machines with pinned dependency versions, and provide all inputs and parameters used to generate their results.
Methods should be implemented as well-documented, reusable components within open-source \ac{HRE} frameworks.
}

\stakeholderProposals{Industry and Funders}{
\termbox{MT}~Industry can accelerate reproducibility by dedicating engineering effort to extending and maintaining public \ac{HRE} tooling.
\termbox{MT}~Funders of academic research should consider supporting multi-year maintenance grants for open-source \ac{HRE} infrastructure that focus on tooling rather than algorithmic novelty.
}

\paragraph{Opportunity II: Ensure Rigorous Comparability.}
Even when authors publish functional and exercisable algorithms, progress is often hard to judge: papers evaluate on incomparable benchmarks, rely on private datasets, and utilize metrics that do not translate across methods.
Meaningful comparison is often prevented by a lack of community-wide benchmarks.
To serve as reliable evaluation targets, future public benchmarks should (i)~be legally cleared, (ii)~provide a ground truth where possible, and (iii)~use agreed-upon metrics tied to practical utility.
Examples include: \ac{IC} images paired with \ac{GDSII} and netlists; \ac{FPGA} bitstreams paired with original netlists; and netlists paired with gate-level labels, module hierarchy, \ac{RTL} description, and design specifications.
Because manufacturing and the supply chain threats addressed by \ac{HRE} evolve rapidly, benchmarks must be treated as living infrastructure with versioning and long-term curation.

\stakeholderProposals{Venues and Community Norms}{
\termbox{ST}~Venues should recognize the creation and maintenance of benchmarks as an integral part of research impact.
Especially for widely cited claims, replication and re-evaluation on contemporary technologies should be encouraged.
Inspiration for organizing such efforts may be drawn from fields such as machine learning, where sustained, community-driven benchmarks have become central to progress~\cite{Hu2020OpenGB, Chao2024JailbreakBenchAO}.
\termbox{MT}~For benchmark creation in \ac{HRE}, initiatives such as ICCAD's annual CAD Contest~\cite{iccad2025cadcontest}, which featured netlist reverse engineering challenges in the past, could serve as a valuable starting point.
Existing efforts like TrustHUB~\cite{trusthub} provide centralized dataset hosting, but broader community engagement and sustained curation appear limited~\cite{DBLP:conf/iccad/Krieg23}, highlighting the need for clear governance, a reliable ground truth, and ongoing maintenance.
}

\stakeholderProposals{Individual Researchers}{
\termbox{ST}~Researchers should provide evaluation scripts and (where applicable) evaluate their tools on public benchmarks using (multiple) suitable, established metrics.
To support reproducibility and reuse, publications should explicitly state assumptions, limitations, and potential failure scenarios.
}

\stakeholderProposals{Industry}{
\termbox{ST}~A realistic starting point is to develop benchmarks from open-source hardware designs with some involvement from companies specialized in \ac{HRE}.
Beyond manufacturers, such firms can contribute reference designs, reproducible ground truth, and representative tasks, and have concrete incentives to do so---visibility and recruitment, and a hand in shaping the methodologies and metrics they themselves rely on.
While such benchmarks would not capture every challenge of state-of-the-art commercial designs, they establish shared methodology and metrics, accelerating \ac{HRE} research.
\termbox{LT}~Benchmarks reflecting state-of-the-art commercial designs---current technology nodes and the design complexity and vendor-specific constraints absent from open-source hardware---may require stronger manufacturer participation, since creating and curating such datasets often exceeds academic capabilities.
The industry of silicon manufacturers, unlike industry specialized in \ac{HRE}, is uniquely positioned to contribute legally cleared commercial designs, verified ground truth, and resources for long-term curation, including updates, governance, and documentation.
We acknowledge that this goal is idealistic, as companies may lack incentives to release such benchmarks.
}

\paragraph{Opportunity III: Provide Legal Clearance and Clarity.}
Legal constraints and uncertainties are structural inhibitors of \ac{HRE} research: they limit the sharing of datasets and tools and lead to vague, insufficient technical descriptions, hampering replication.
Constraints arise from licensing terms, non-disclosure agreements, and export-control regulations.
For instance, open-source \ac{IC} reverse-engineering tools could be classified as dual-use technologies under \ac{EU} export regulations, requiring explicit export approval~\cite{EU2021_821,EU2025_2003}.
\ac{IP} concerns introduce further uncertainty about which results can be published.
We argue that such restrictions are increasingly misaligned with contemporary hardware security needs.
For instance, this misalignment manifests in workforce development: unlike \ac{SRE}, \ac{HRE} lacks a broad talent pipeline due to the high barrier to entry~\cite{DBLP:conf/sigcse/WalendyW0PR25}.
Public research and education are among the few viable entry points for training practitioners, and they require appropriate legal protection.

\stakeholderProposals{Government and Regulators}{
\termbox{LT}~Governments should reassess legal barriers and export restrictions that disproportionately impede \ac{HRE}, support broad access to publicly funded research outputs, and fund public-private mechanisms for lawful sharing and long-term maintenance.
The \ac{SRE} ecosystem shows that government-backed open-source tools can flourish despite dual-use concerns (\eg, Ghidra).
We recognize that governments with strong domestic semiconductor industries may be reluctant to relax export controls, particularly for leading-edge technology.
We do not argue for wholesale deregulation, but for carving out space for public research and education: the alternative---capability concentrated in a few cleared actors---erodes the independent verification on which trustworthy hardware ultimately depends.
}

\stakeholderProposals{Industry}{
\termbox{MT}~Industry can reduce friction without compromising \ac{IP} by offering research-friendly licensing terms for benchmark use and publication, at least for selected, non-critical hardware designs and implementations.
After all, legal restrictions often bind only law-abiding researchers, while adversaries ignore them.
}

\stakeholderProposals{Venues and Institutions}{
\termbox{ST}~To reduce legal uncertainty for researchers, venues and institutions should provide clear guidance on legally constrained artifacts, including sanitization techniques and minimum disclosure requirements.
\termbox{MT}~Well-resourced institutions could further support researchers through legal consultation, handling approvals, or negotiating exemptions, enabling broader participation in \ac{HRE} research.
}

\section{Conclusion}
\label{section:conclusion}
In this work, we identify both domain-specific and cross-cutting ways forward to strengthen the future of \ac{HRE} research.
Specifically, three structural barriers underlie a systemic crisis in this field: scarce reusable artifacts, missing benchmarks, and unresolved legal constraints---barriers that span \ac{IC}, \ac{FPGA}, and netlist reverse engineering and require coordinated, stakeholder-specific action from academia, government, and industry.
Together, these barriers have produced a research ecosystem that has made substantial progress on individual subproblems but has not yet realized the opportunities enabled by reproducible, cumulative science---a finding underscored by a retrospective artifact evaluation in which key results could be reproduced for only 4\% of publications. 
In addition to describing opportunities to address these systemic issues, we identify eleven concrete technical challenges and propose actionable directions for each. 
We base these findings on an extensive systematic review of 187 publications spanning more than two decades of \ac{HRE} research.
Our work calls for a transition from isolated research silos toward a collaborative research discipline capable of assuring increasingly complex hardware supply chains.

\ifanonymous
\else
\section*{Acknowledgments}
We thank Jonas Schmitt and Rudolfs Pogodins for their support with the artifact evaluation.
This work was supported by the Deutsche Forschungsgemeinschaft (DFG, German Research Foundation) under Germany’s Excellence Strategy -- EXC 2092 CASA -- 390781972, and by the \href{https://rc-trust.ai}{Research Center Trustworthy Data Science and Security}, one of the Research Alliance Centers within the \href{https://uaruhr.de}{UA Ruhr}.

\paragraph{CRediT Authorship Contribution Statement.}

\textbf{Methodology}: SB, JS, RW.
\textbf{Software}: KD, RW.
\textbf{Investigation}: FH, ZK, SK, JS, RW.
\textbf{Data Curation}: SB, FH, ZK, SK, JS, RW.
\textbf{Writing -- Original Draft}: SB, FH, ZK, SK, JS, RW.
\textbf{Writing -- Review \& Editing}:  SB, FH, ZK, SK, JS, RW.
\textbf{Visualization}: KD, ZK, JS.
\textbf{Supervision}:  SB, CP.
\textbf{Project Administration}: ZK, RW.
\textbf{Funding Acquisition}: SB, CP.

\fi

\bibliographystyle{alpha}
\bibliography{bibliography, bibliograpghy_asic_fpga_netlist}


\end{document}